\documentclass[12pt]{article}
\usepackage{float}
\usepackage{pgf, tikz}
\usetikzlibrary{arrows, automata}

\usepackage[textwidth=8em,textsize=small]{todonotes}
\usepackage[utf8]{inputenc}
\usepackage{mathbbol}
\usepackage{bm}
\usepackage{url}
\usepackage{amsmath}
\usepackage{natbib}
\usepackage[utf8]{inputenc}
\usepackage{xcolor}
\usepackage{graphicx}
\usepackage{enumerate}
\usepackage{amsthm}
\usepackage{amsmath}
\usepackage{amssymb}
\usepackage{marginnote}
\usepackage{mathtools}
\usepackage{float}

\usepackage{booktabs} 
\usepackage{changepage}
\usepackage{color}
\usepackage{graphicx}
\usepackage{setspace}
\usepackage{lscape}
\usepackage{algorithm}
\usepackage{algpseudocode}

\usepackage{etoolbox}
\makeatletter
\patchcmd{\@makecaption}
  {\parbox}
  {\advance\@tempdima-\fontdimen2} 
  {}{}
\makeatother  

\usepackage[top=0.7in, bottom=0.7in, left=1in, right=1in]{geometry}

\begin{document}

\begin{center}
{\Large Semiparametric Inference of Effective Reproduction Number Dynamics from Wastewater Pathogen Surveillance Data 
}\\ \ \\

Isaac H. Goldstein$^1$, 
Daniel M. Parker$^2$, 
Sunny Jiang$^3$,
Volodymyr M. Minin$^1$\\
$^1$Department of Statistics, University of California, Irvine \\
$^2$Department of Population Health and Disease Prevention; Department of Epidemiology and Biostatistics, University of California, Irvine \\
$^3$Department of Civil and Environmental Engineering; Department of Ecology and Evolutionary Biology, University of California, Irvine\\[2pt]
\end{center}



\begin{abstract}{
Concentrations of pathogen genomes measured in wastewater have recently become available as a new data source to use when modeling the spread of infectious diseases. 
One promising use for this data source is inference of the effective reproduction number, the average number of individuals a newly infected person will infect. 
We propose a model where new infections arrive according to a time-varying immigration rate which can be interpreted as an average number of secondary infections produced by one infectious individual per unit time.
This model allows us to estimate the effective reproduction number from concentrations of pathogen genomes while avoiding difficult to verify assumptions about the dynamics of the susceptible population. 
As a byproduct of our primary goal, we also produce a new model for estimating the effective reproduction number from case data using the same framework.  
We test this modeling framework in an agent-based simulation study with a realistic data generating mechanism which accounts for the time-varying dynamics of pathogen shedding. 
Finally, we apply our new model to estimating the effective reproduction number of SARS-CoV-2 in Los Angeles, California, using pathogen RNA concentrations collected from a large wastewater treatment facility. 
}
\end{abstract}


\section{Introduction}
\label{sec:intro}
For many pathogens, infected individuals will shed copies of the pathogen through fecal matter over the course of their infection. 
Viral gene concentrations measured in wastewater samples are a noisy aggregate of the concentrations of genomes generated by infected individuals connected to the wastewater system, and thus provide insight into the dynamics of the spread of an infectious disease \citep{hillary2020wastewater,polo2020making}. 
One promising use for pathogen genome concentrations as data is estimation of the effective reproduction number. 
The effective reproduction number ($R_{t}$), the average number of individuals an infectious person at time $t$ would subsequently infect, is a useful way of describing the state of an infectious disease epidemic. 
When $R_{t}$ is below 1, we expect the number of new infections to decrease; the reverse is true when $R_{t}$ is above 1. 
In this paper, we develop a new method for estimating the effective reproduction number from pathogen genome concentrations collected from wastewater, and as a bi-product, show how this method can be used to estimate the effective reproduction number from case data as well.
\par
Recently, a number of studies have evaluated SARS-CoV-2 (the causative agent of COVID-19) RNA concentrations measured in wastewater as a potential data source, comparing them to both prevalence counts, counts of reported cases, and case rates. \citep{morvan2022analysis,acer2022quantifying,song2021detection,zhan2022relationships,zulli2022predicting}.  
\citet{wade2022understanding} provide a useful introduction to the many sources of uncertainty in the pathogen genome concentration data generation process. 
\par 
From the perspective of inferential methods, pathogen genome concentrations are potentially less biased data than counts of new cases, which can be biased by policies regarding testing availability and the willingness of the population to test \citep{li2020spatial}. 
To our knowledge, there have been relatively few attempts to use pathogen genome concentrations collected from wastewater (henceforth referred to as wastewater data) to estimate the effective reproduction number. 
\citet{huisman2022wastewater} adapted their case based method \citep{huisman2022estimation} and used SARS-CoV-2 wastewater data to create a synthetic time series of hypothetical case data, which is then analyzed with the widely used case-based method \texttt{EpiEstim}  \citep{cori_new_2013}. 
While easy to use, the synthetic incidence is truncated on the assumption that wastewater data observed in the present contains little information about the number of newly infected individuals. As a consequence, the final estimate of the effective reproduction number is truncated as well. 
\citet{nourbakhsh2022wastewater} used a classic compartmental model where RNA concentrations were modeled as noisy realizations of the number of currently infectious and recently recovered individuals. 
While the model produces inference on a number of parameters beyond the effective reproduction number, it also requires a number of parameters to be specified by users in order to produce inference, many of which are difficult to verify in practice. 
\par
Taking inspiration from previous work on non-parametric modeling of the transmission rate in compartmental models \citep{xu2016bayesian} and birth-death modeling in infectious disease phylodynamics \citep{stadler2013birth}, we introduce a simpler compartmental model with only compartments needed to estimate the effective reproduction number and equip it with a Bayesian nonparametric inference framework. 
This simpler model combined with our Bayesian nonparametric framework lets us avoid some difficult-to-verify assumptions, while still estimating the effective reproduction number from wastewater data. 
\par 
In this paper, we first introduce the classic compartmental modeling framework, then our new wastewater-based method for estimating the effective reproduction number. 
We test our new model against compartmental models fit to case and wastewater data as well as a state-of-the-art wastewater-based method on simulated data. 
Finally, we apply our new method to estimate the effective reproduction number of SARS-CoV-2 in Los Angeles, California using SARS-CoV-2 RNA concentrations collected from a large wastewater treatment plant. 

\section{Methods}
\label{sec:methods}
\subsection{Available data}
We will consider two types of surveillance data; concentrations of pathogen genomes measured in wastewater, and reported new cases, observed at times $t_{1}, \dots t_{T}$.
It is common practice to measure the concentration from the same sample of wastewater multiple times, producing multiple measurements called replicates. 
In real world data sets, an average of replicates is often reported. 
We will consider both raw concentrations and averages in this study. 
We define $\mathbf{X} = (X_{t_{1},1}, \dots, X_{t_{1},j}, \dots, X_{t_{T},j})$, where $X_{t_{i},j}$ is the $jth$ replicate of pathogen genomes collected from wastewater at time $t_{i}$, with units of copies per milliliter. 
We model $X_{t_{i},j}$ as a noisy representation of the unobserved number of currently infectious and recently recovered individuals.
Let $\mathbf{O} = (O_{t_{1}}, O_{t_{2}}, O_{t_{3}}, \dots, O_{t_{T}})$, where $O_{t_{u}}$ is the number of newly observed cases of an infectious disease during time interval $(t_{u-1}, t_{u}]$. We model $O_{t_{u}}$ as a noisy realization of the number of individuals transitioning from the latent stage of infection to the infectious stage. 

\subsection{Standard compartmental models}
An SEIR compartmental model describes a homogeneously mixing population moving through infectious disease stages, referred to as compartments \citep[pages 12-52]{keeling2008modeling}. 
The compartments are $S$, susceptible individuals; $E$, infected but not yet infectious individuals; $I$, currently infectious individuals; and $R$, no longer infectious either due to recovery or death. 
In its deterministic form, the changes in the number of individuals in these compartments are described using a system of ordinary differential equations (ODEs). 
The behavior of the ODEs is described by a set of key parameters defined in Table \ref{tbl:ode_param}.
\begin{table}
    \centering
         \caption{Parameters of the SEIR model.}
    \begin{tabular}{cl}
        Parameter & Interpretation  \\
        \hline
        $\beta_{t}$ & time-varying transmission rate\\
        $1/\gamma$ & average time infected but not infectious (average length of the latent period)\\
        $1/\nu$ & average length of the infectious period\\
        $N$ & total population size
    \end{tabular}
    \label{tbl:ode_param}
\end{table}
The SEIR system of ODEs is:
\begin{equation*}
\frac{dS}{dt} = -\beta_{t}\times I \times S/N, \quad
\frac{dE}{dt} = \beta_{t}\times I \times S/N - \gamma \times E, \quad
\frac{dI}{dt} = \gamma \times E - \nu \times I, \quad
\frac{dR}{dt} = \nu \times I.
\end{equation*}
We model $\beta_{t}$ as time-varying to account for changes in transmission due to, for example, implementation of public health policies, changes in behavior, or the emergence of new pathogen variants.
\par 
For the SEIR model, the time-varying basic reproduction number, $R_{0,t}$, the average number of individuals a person infected at time $t$ would infect in a completely susceptible population, and effective reproduction number, $R_{t}$, are defined as: 
\begin{equation}\label{eqn:rt_def}
R_{0,t} = \frac{\beta_{t}}{\nu}, \; R_{t} =  R_{0,t} \times \frac{S(t)}{N}.
\end{equation}
We will adapt this classic model for our purpose of estimating $R_t$ from wastewater data.

\subsection{The EIRR model}\label{eirr_model}
The SEIR model assumes that the susceptible population only changes because of new infections. 
In practice, the susceptible population can change over time due to vaccination campaigns and the introduction of new disease variants that evade immunity from prior infection.
Modeling such dynamics is often difficult, especially in real time, when it may be non-trivial to estimate the effect of immune evasion of a particular variant.

\par 
Taking inspiration from birth-death modeling in infectious disease phylodynamics \citep{stadler2013birth}, we define $\alpha_{t} = \beta_{t} \times S/N$ and rewrite the equation for $dE/dt$ so that: $dE/dt = \alpha_{t} \times I - \gamma \times E.$
The rate of new latent infections no longer depends on the $S$ compartment. 
The parameter $\alpha_{t}$ can be interpreted as a time-varying average number of secondary infections produced by one infectious individual per unit time (e.g., per day).
Note, the effective reproduction number is still recoverable, as 
$R_{t} = \beta_{t} S(t)/\nu N = \alpha_{t}/\nu$.
\par 
In addition, we split the $R$ compartment in two. 
In the first compartment individuals are recovered but still shedding pathogen genomes, in the second they are recovered and no longer shedding genomes. 
This choice is motivated by the characteristics of SARS-CoV-2, for which it has been shown individuals shed detectable amounts of RNA in fecal matter well after the likely end of their infectious period (see Web Section A.2.3) \citep{okita2022duration,zhang2021prevalence}. 
The final model, which we call the EIRR model, is described by the following equations:
\begin{equation}\label{eqn:eirr}
\frac{dE}{dt} = \alpha_{t} \times I - \gamma \times E,\quad
\frac{dI}{dt} = \gamma \times E - \nu \times I, \quad
\frac{dR1}{dt} = \nu \times I - \eta \times R1,\quad
\frac{dR2}{dt} = \eta \times R1.
\end{equation}
Here $1/\eta$ is loosely interpreted as the average time spent recovered but still shedding pathogen RNA via fecal matter.
We also include a redundant compartment $C(t)$, where $dC/dt = \gamma \times E.$ This counts cumulative transitions from the $E$ to $I$ compartments, and allows us to keep track of the number of people who became infectious during time period $(t_u, t_{u-1}]$ as $C(t_{u})-C(t_{u-1})$. 
For the sake of comparison, we will also implement the SEIRR model, which is the SEIR model with two R compartments. 

\subsection{Wastewater observation model}\label{obs_model}
We model the log of observed pathogen genome concentrations as realizations of a generalized t-distribution: $\log{X_{t_i,j}} \sim \text{Generalized t}(\log{\left(\lambda \times I(t_{i}) + (1-\lambda) \times R1(t_{i})\right)} + \log{(\rho)}, \tau^{2}, df)$. Here $I(t_{i})$ is the number of currently infectious individuals at time $t_{i}$, $R1(t_{i})$ is the number of non-infectious but still shedding individuals at time $t_{i}$.
The parameter $\lambda \in (0,1)$ is a normalized weight representing how much each individual contributes to the true underlying pathogen genome concentrations while infectious (Web Section A.1.1 describes in detail how we chose the prior for this parameter).
Parameter $\rho$ allows for flexibility in relating counts of individuals to concentrations.
Parameter $\tau$ accounts for variation from the mean, and $df$ is the parameter governing the degrees of freedom of the t-distribution.
We chose to use a t-distribution because wastewater data often has many outliers; a t-distribution with thicker tails should better fit the data as opposed to the normal distribution.

\subsection{Complete EIRR-ww model structure}
We describe the complete wastewater data model, which we call the EIRR-ww model, structure in the following section. 
We use a random walk prior for the time-varying effective reproduction number: $R_{0} \sim \text{Log-Normal}(\mu_{0}, \sigma_{0}), \sigma \sim \text{Log-Normal}(\mu_{rw}, \sigma_{rw}), \log{(R_{k_{i}})}|R_{k_{i-1}}, \sigma \sim \text{Normal}(\log{(R_{k_{i-1}})}, \sigma)$.
The times $k_{i}$ can be chosen flexibly, for this study, we choose them so that $R_{t}$ changes on a weekly basis. 
Let $\mathbf{\bar{\Theta}} = (\gamma, \nu,\eta, I(0), E(0), R1(0))$
and $\mathbf{R} = (R_{k_{1}}, \dots, R_{k_{M}})$ be the vector of effective reproduction number values. 
Let $M( t, \mathbf{\bar{\Theta}}, \mathbf{R})$ $=$\\ $(\textbf{E}(t, \mathbf{\bar{\Theta}}, \mathbf{R}), \textbf{I}(t, \mathbf{\bar{\Theta}}, \mathbf{R}), \textbf{R1}(t, \mathbf{\bar{\Theta}}, \mathbf{R}), \textbf{R2}(t, \mathbf{\bar{\Theta}}, \mathbf{R}))$ 
be the solution to the EIRR ODE system described in Section \ref{eirr_model}. 
The target posterior distribution is: 
\begin{equation*}
        P(\textbf{R}, \mathbf{\bar{\Theta}}, \rho, \lambda, \tau, df, \sigma \mid \textbf{X}) \propto
                \mathrlap{\underbrace{\phantom{P(\textbf{X} \mid \textbf{M}(t, \mathbf{\bar{\Theta}}, \mathbf{R}), \rho, \lambda, \tau, df)}}_{\text{Concentration Model}}}
P(\textbf{X} \mid \textbf{M}(t, \mathbf{\bar{\Theta}}, \mathbf{R}),  \rho, \lambda, \tau, df)
 \mathrlap{\underbrace{\phantom{P(\mathbf{R} \mid \sigma)}}_{\text{RW Prior}}}P(\mathbf{R} \mid \sigma)P(\mathbf{\bar{\Theta}}, \rho, \lambda, \tau, df, \sigma).
\end{equation*}
We use the No-U-Turn Sampler, implemented in the \texttt{Julia} package \texttt{Turing} to approximate this posterior distribution \citep{NUTS,turing}.
We used non-centered re-parameterizations for all model parameters except for $df$ (which had a gamma prior).
Markov chain Monte Carlo chains were initialized using the Maximum A Posterior (MAP) estimate of each parameter plus Gaussian noise (except for $df$ which was only initialized at the MAP).

\subsection{The EIR model}
The EIRR model can be simplified when fitting to case data by using only a single $R$ compartment, creating the EIR model. 
It is described by the same equations as Equation \ref{eqn:eirr} but with only one $R$ compartment equation.
Cases are modeled as a noisy realization of the number of transitions from the E to the I compartment using a negative-binomial likelihood. For cases observed in the interval $(t_{u-1}, t_{u}]$: $O_{u} \sim \text{Negative-Binomial}((C(t_{u}) - C(t_{u-1})) \times \psi, \phi)$
where $C(t_{u}) - C(t_{u-1})$ is the number of transitions from the $E$ to the $I$ compartment in time interval $(t_{u-1}, t_{u}]$, $\psi$ is a detection rate parameter, and $\phi$ is an over-dispersion parameter. 
Both $\psi$ and $\phi$ have their own priors.
The full structure of our case model, the EIR-cases model, is otherwise very similar to the EIRR-ww model.
The structures of the corresponding SEIR-cases/SEIRR-ww models are likewise similar, though for SEIR-cases/SEIRR-ww models the basic reproduction number is modeled as random walk, rather than the effective reproduction number. 
We provide a more detailed description of the SEIRR-ww model priors in Web Section A.1.3. 
\par 
All code used to produce this paper is available at \url{https://github.com/igoldsteinh/ww_paper}. 
A \texttt{Julia} package implementing the models used in this paper is available at \url{https://github.com/igoldsteinh/concRt.jl}. 
An \texttt{R} package which provides a wrapper for the \text{Julia} package is available at \url{https://github.com/igoldsteinh/concRt}. 

\section{Simulation}
\subsection{Simulation Protocols}
We simulated a single realization from an agent-based stochastic SEIRR described in Web Sections  A.1.1 and A.2.1. 
The population size was set to 100,000. 
The mean latent period was 4 days, the mean infectious period was 7 days, and the mean time spent recovered but still shedding pathogen genomes was 18 days.
The simulation was started with 200 individuals in each of the $E$ and $I$ compartments, and run for a warm-up period of 77 days before creating data for the model to fit to. 
This was done so that there would be individuals in all compartments for whom all transition times between compartments would be naturally available. 
The top left panel of Figure A8 shows the prevalence in each compartment, the time 0 is the day before the first wastewater sample is collected. 
The basic reproduction number $R_{0,t}$ was given a fixed trajectory. We calculated the true $R_{t}$ at each day using Equation \ref{eqn:rt_def}. 
For the observation period, we chose to start with $R_{0,t}$ set to $0.9$ with a rapid increase to $2.5$, where it stayed for the duration of the simulation, mimicking a scenario where a new and highly infectious variant is introduced into a population. 
All priors used in the simulation are listed in Table A2 (for the case models the priors for $\gamma$ and $\nu$ are transformed to be on a weekly scale).
Note that the prior for $\lambda$ was centered at $0.99$, with a 95\% quantile range of 0.8 to 1.
\par
Using this single realization from the stochastic SEIRR model, we simulated 100 data sets of pathogen genome concentrations and 100 data sets of observed case data. 
All parameters specified below were chosen to create data similar to observed data from the SARS-CoV-2 pandemic in Los Angeles, California (
see Web Section A.2.3 for more details).
Daily genome concentration data was generated using a generalized t-distribution as in the model described in Section \ref{obs_model}. 
However, the mean of the generalized t-distribution was the true total genome concentration shed, generated using the method described in Web Supplementary Material Section A.1.1. 
We simulated ten replicates per day. 
Parameter $\rho$ was set to be $0.011$, $\tau$ was set to be $0.5$, and $df$ was set to $2.99$.
Only data from every other day were used for analysis.
Cases were simulated at a daily time-scale and aggregated to a weekly time-scale. 
The case detection rate $\psi$ was set to $0.2$, while $\phi$ was set to $57.55$.
We generated data for a total of 19 weeks.
While the case likelihood of our SEIR-cases and EIR-cases models is quite similar to the data generating mechanism of the simulated data, the wastewater likelihood of our SEIRR-ww and EIRR-ww models is a crude approximation of the data generating mechanism of the simulated data.
For all subsequent simulation scenarios except the final one, we use the same 100 data sets while changing either the type of data used to fit the model, or the model priors.
\par 
In the baseline simulation scenario, we fit the SEIR-cases, EIR-cases, SEIRR-ww and EIRR-ww models to the data, using three replicates for the wastewater models and weekly cases for the case models. 
In subsequent scenarios, we only fit the EIRR-ww model. 
Using the same priors as in the baseline scenario, we fit the EIRR-ww model using one or ten replicates instead of three replicates (1-rep and 10-rep) and also fit the model using the mean of three replicates or the mean of ten replicates (3-mean and 10-mean respectively). 
We also conduct sensitivity analyses where the priors for the inital E and I compartments are centered at 75\% or 133\% of true values (Low Init and High Init respectively). 
We shift the prior for $\lambda$ so that it is centered around $0.8$ (Low Prop) as opposed to the default $0.99$. 
We fit the Huisman method using the mean of three replicates as the input data. 
Details on choosing parameters for the Huisman method are in Web Section A.1.6.
In the final scenario (Stoch $R_{t}$), we use the parameters, data, and priors of the baseline scenario, but for each data set simulate a new epidemic, and thus a new $R_{t}$ curve, for each simulation.
An example realization of the simulation is displayed in Figure A8.
All priors used in the baseline simulation are listed in Table A2.

\subsection{Comparison with state-of-the-art methods}
We compare the EIRR-ww model to the \citet{huisman2022wastewater} method.
This method is a variation on the well known \texttt{EpiEstim} method \citep{cori_new_2013}. 
Pathogen concentrations are modeled as a convolution of unobserved latent incidence (new infections) and the individual shedding load profile describing how many gene copies individuals shed over the course of their infection. 
Latent incidence is estimated using an EM algorithm, then the estimated incidence is used as the input into \texttt{EpiEstim}. 
The pipeline is repeated multiple times using a bootstrap method to produce final measures of uncertainty. 
Further details are available in Web Section A.1.2.
\par 
In contrast, the method of \citet{nourbakhsh2022wastewater} uses a compartmental model similar to ours, splitting both the $I$ and $R1$ compartments into many smaller compartments in order to better match the shedding dynamics of pathogen genomes in fecal matter. 
They also directly model the impact of the sewer system itself on the final observed data.
We decided not to test the method of \citet{nourbakhsh2022wastewater} in this study because the code for the latter method is not readily available, and because the goal of their model was not limited to inference of the effective reproduction number.

\subsection{Simulation results}
\subsubsection{Baseline simulation}
\begin{figure}
    \centering
    \includegraphics[width = 1.0\textwidth]{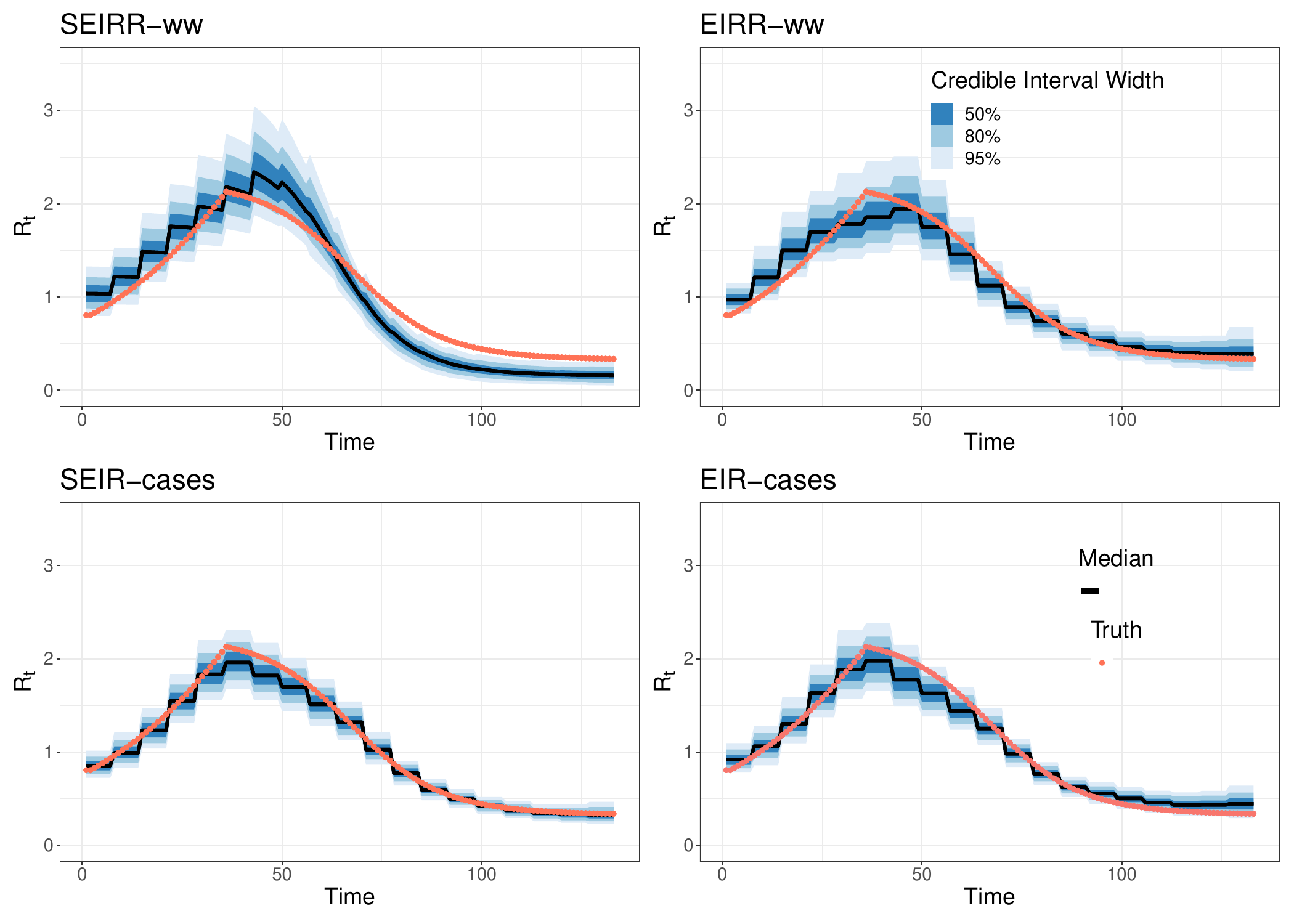}
    \caption{Posterior summaries of Rt using four models fit to either wastewater or case data generated from the same underlying infectious disease dynamics. True Rt trajectories are colored in red, black lines represent posterior medians, blue shaded areas from dark to light are 50, 80 and 95\% credible intervals. Models in the top row use genome concentrations, models in the bottom row use case counts. Models in the left column use the $S$ compartment, models in the right column do not.}
    \label{fig:example_posteriors}
\end{figure}

Posterior medians and credible intervals for models fit to the example simulation data are displayed in Figure \ref{fig:example_posteriors}. 

Posterior trajectories from all models generally mimic the true $R_{t}$ curve, although the SEIRR-ww model struggles to capture the exact trajectory. 
Models using genome concentrations have wider credible intervals than models using case counts, reflecting higher variability of wastewater data as compared to case data. 
Note that the random walk prior forces $R_{t}$ to change on a weekly scale, while the true values are reported on a daily scale, resulting in more pronounced segmentation of posterior summaries. 
The SEIRR-ww model (top left of Figure \ref{fig:example_posteriors}) estimates sloping spikes because the SEIRR ODEs are solved at a daily time scale, since the SEIR-cases model is solved at a weekly scale, it does not display the same behavior. 
EIRR-ww posterior estimates of the latent trajectories (including latent incidence) are displayed in Web Figure A9. 
While posterior estimates mimic the shape of the latent trajectories, they fail to capture the magnitude of the trajectories except at the beginning and end of the simulation.

\begin{figure}
    \centering
    \includegraphics[width = 1.0\textwidth]{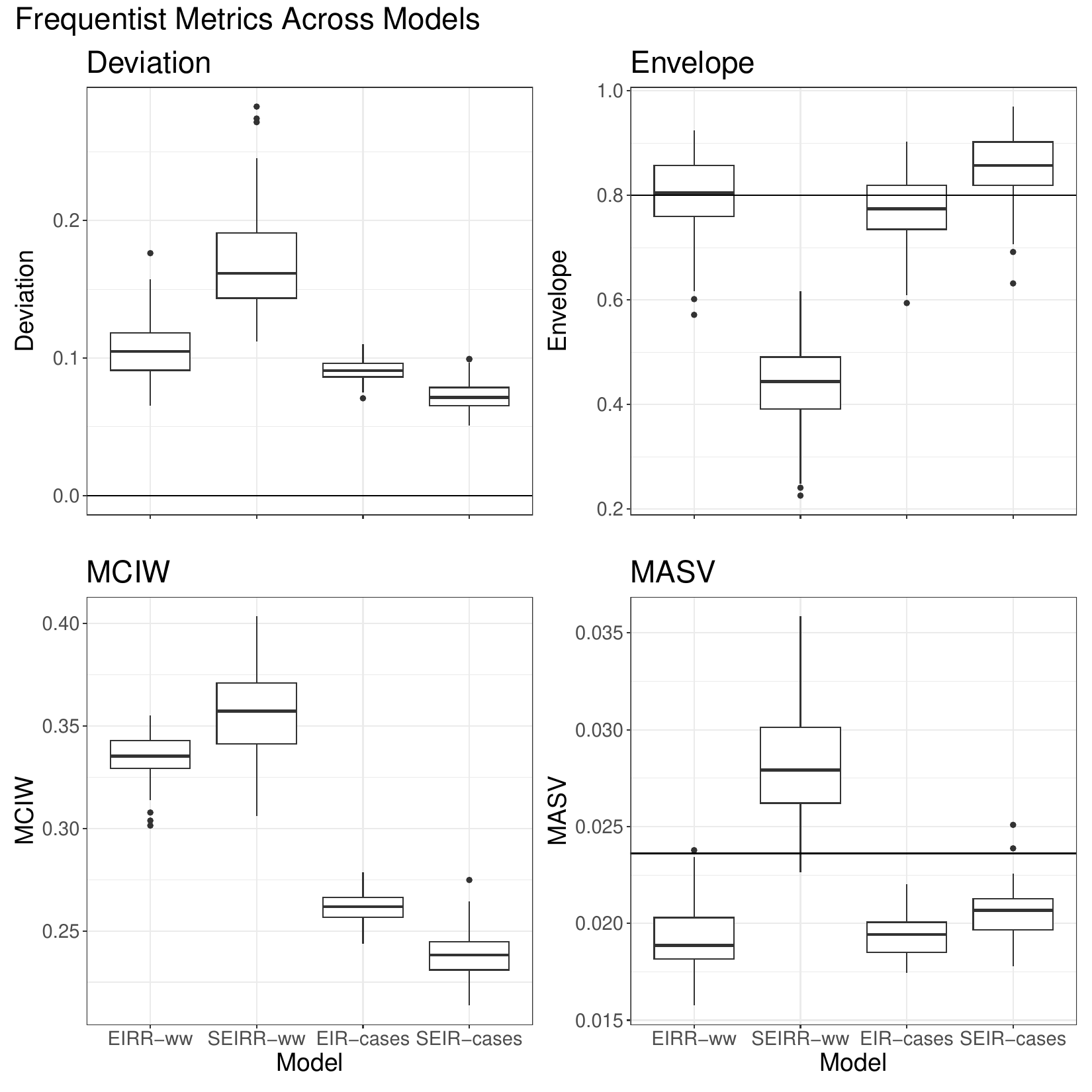}
    \caption{Frequentist metrics for the EIRR-ww, SEIRR-ww, EIR-cases and SEIR-cases models in the baseline scenario. Absolute deviation is the mean of the absolute value of the difference between the median $R_{t}$ at each time point and the true value. Envelope is a measure of coverage, taking the average coverage of 80\% intervals over the time series. MCIW is the average mean credible interval width. Mean absolute standard deviation (MASV) is the difference between the current median point estimate for $R_{t}$ and the previous point estimate for $R_{t}$. The line in the bottom right panel represents the true absolute standard deviation. Boxplot solid lines represent medians, hinges are upper and lower quartiles and whiskers are at most 1.5 times larger than the upper and lower quartiles.}
    \label{fig:freq_metrics_baseline}
\end{figure}

To assess performance across many simulated data sets, we examined frequentist properties of our four models, summarized in Figure \ref{fig:freq_metrics_baseline}. 
Boxplot solid lines represent medians, hinges are upper and lower quartiles and whiskers are at most 1.5 times larger than the upper and lower quartiles. 
Envelope is a measure of coverage. 
For each simulation the envelope is the proportion of time points for which an 80\% credible interval from the posterior distribution captured the true value of interest. 
Ideally it should be 0.8. 
We chose to use 80\% credible intervals as estimates of the 80\% quantiles have less Monte Carlo Error than 95\% quantiles, so fewer data sets are needed to estimate them well. 
The corresponding 95\% credible interval results are displayed in Web Figure A13. 
Mean credible interval width (MCIW) is the mean of 80\% credible interval widths across time points within a simulation. 
Absolute deviation is a measure of bias, and is the mean of the absolute difference between the posterior median and the true value at each time point. 
Finally, mean absolute sequential variation (MASV) measures how well each method captures the variation in the effective reproduction number across time by computing the mean of the absolute difference between the posterior median at $t$ and the posterior median at $t-1$. 
We compare this to the true mean absolute sequential variation in each simulation.
The EIRR-ww model outperforms the SEIRR-ww model in terms of bias, precision, and coverage. 
Both the SEIR-cases model and the EIR-cases model outperform the EIRR-ww model in terms of bias and precision. 
The EIR-cases model is slightly more biased and less precise than the SEIR-cases model. 
Although the EIRR-ww model has less precision than case based models, this simulation shows that models using wastewater data can be used to estimate the effective reproduction number reasonably well. 

\subsubsection{Performance in other scenarios}
Frequentist metrics comparing the baseline EIRR-ww fit (three replicates) to the EIRR-ww fit to one and ten replicates, and to the EIRR-ww fit to mean of three replicates and mean of ten replicates, are displayed in Web Figure A14. 
The EIRR-ww fit to one or ten replicates performed modestly worse or better respectively than the EIRR-ww fit to three replicates.
The EIRR-ww fit to raw replicate concentrations performed slightly better than the EIRR-ww fit to means of replicates.
\par
We assessed the robustness of our model by changing the priors for the initial conditions, as well as the prior for $\lambda$. 
Frequentist metrics comparing these alternate models to the baseline model are displayed in Figure A15. 
Changing these priors lead to only modest changes in model performance. 
\par
We fit the method by \citet{huisman2022wastewater} to each of our data sets, using as the input data the mean of three replicates. 
We then compared the \citet{huisman2022wastewater} method to the EIRR-ww model fit to three replicates. 
Also, the Huisman et al. method does not provide 80\% credible intervals, so we compared metrics using 95\% credible intervals. 
The comparison is visualized in Figure A16. 
In this simulation, the EIRR-ww model clearly outperforms the Huisman model in terms of both bias and precision associated with the effective reproduction number trajectory estimation. 
We found that the EIRR-ww model performed similarly to the baseline scenario when fit to 100 data sets where each data set was generated from a separate simulated epidemic and separate simulated $R_{t}$.
The comparison is visualized in Figure A17.
Summaries of MCMC diagnostics for all model fits are available in Web Section A.2.8.

\section{The effective reproduction number of SARS-CoV-2 in Los Angeles, CA}
\begin{figure}
    \centering
    \includegraphics[width = 1.0\textwidth]{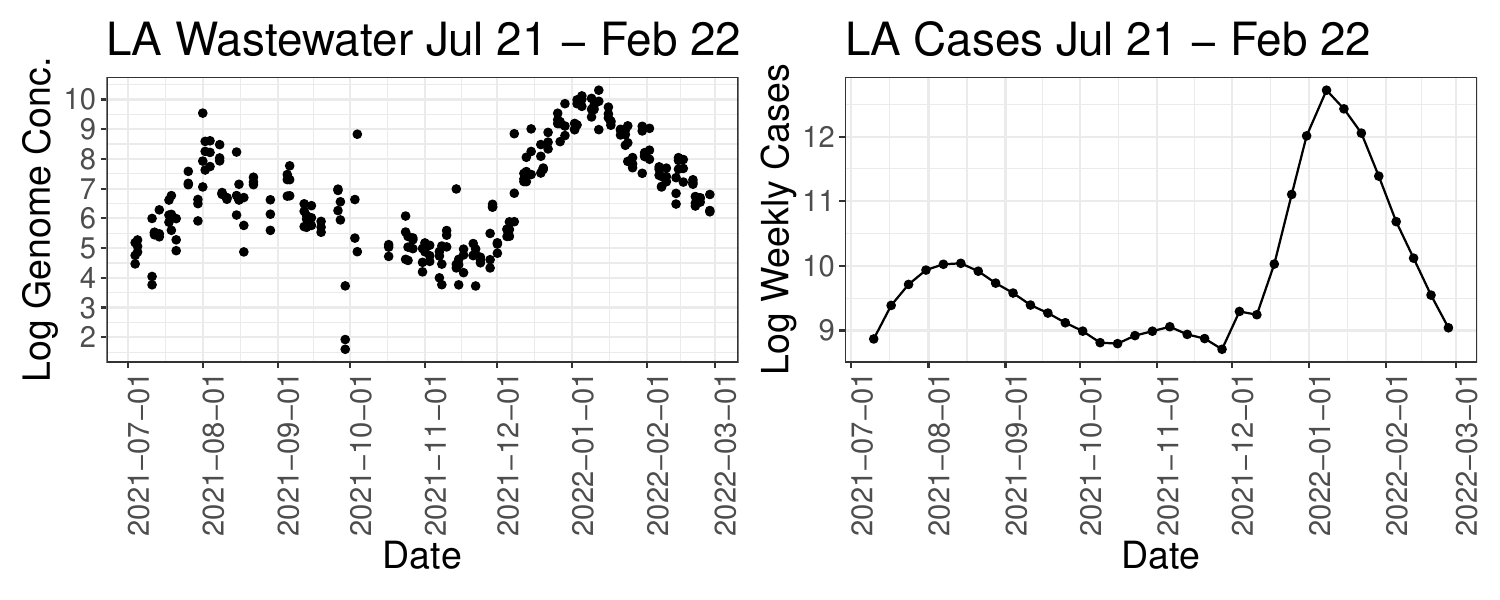}
    \caption{Wastewater and case data for the SARS-CoV-2 epidemic in Los Angeles, CA.}
    \label{fig:la_data}
\end{figure}

\begin{figure}
    \centering
    \includegraphics[width = 1.0\textwidth]{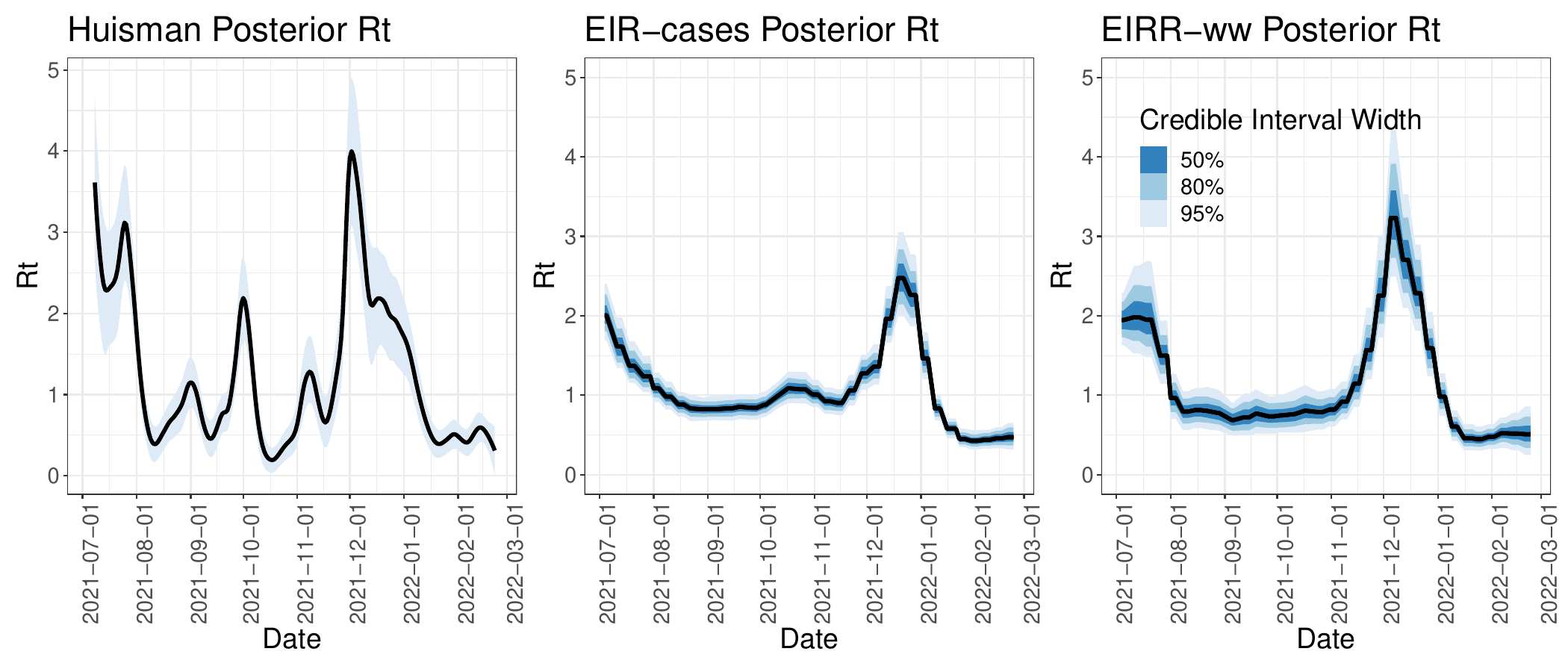}
    \caption{Posterior summaries of the effective reproduction number for the SARS-CoV-2 Delta and Omicron waves in Los Angeles, CA. Blue bars from dark to light represent 50, 80, and 95\% credible intervals. Black lines represent median posterior estimates. EIRR-ww model and Huisman model are fit to RNA concentrations collected from wastewater data, EIR-cases model is fit to weekly case counts.}
    \label{fig:la_rt}
\end{figure}

Wastewater data was collected from the Joint Water Pollution Control Plant (JWPCP), one of the largest wastewater treatment plants in Los Angeles County. 
The plant serves 4.8 million people across Los Angeles County. 
The data, reported as viral gene copies per ml of wastewater determined by quantitative PCR, were collected from the 24-hour composite wastewater influent samples at irregular, but approximately two-day intervals, and usually three replicates were reported for each sample \citep{song2021detection}. 
We excluded two days (7/7/21 and 8/23/21) as outliers, as the reported concentrations dropped by at least two orders of magnitude compared to the concentrations of the closest previous and subsequent observed days.
Cases during the same period in Los Angeles County are available from the California Open Data Portal \citep{caopendata}. 
The available data from cases and wastewater are visualized in Figure \ref{fig:la_data}.
The cases are recorded for all of Los Angeles County, not just the population served by the JWPCP plant.
We re-scaled the cases by a factor of 0.48 to partially account for this, as there are about 10 million people in Los Angeles County in total. 
Priors for the EIRR-ww and EIR-cases model were the same as those used in the simulation, except for the priors on the initial conditions and initial $R_{t}$ (see Web Section A.2.10). 
For the \citet{huisman2022wastewater} method, we used the same shedding load profile calculated for the simulated data sets, but used the mean and and standard deviation of the generation time distribution of SARS-CoV-2 calculated by \citet{Sender2021}. 
The posterior estimate of $R_{t}$ (from left to right) of the \citet{huisman2022wastewater} method, the EIR-cases model and the EIRR-ww model are shown in Figure \ref{fig:la_rt}. 
For additional comparisons, we used two branching process models, the Rt-estim-gamma model fit to cases and total number of diagnostic tests \citep{goldstein2022incorporating} and another model fit to cases using the \texttt{epidemia} package \citep{epidemia_paper,epidemia}. 
For more details on these models, see Web Section A.1.7. Posteriors from these models, along with the EIR-cases and EIRR-ww models are shown in in Figure A18.
\par 
While the EIRR-ww model provides different estimates than any of the case-based methods, they mostly align, estimating one large increase in $R_{t}$ above 1 tied with the arrival of the Omicron variant in California in winter 2021. 
In contrast, the Huisman method estimates several dramatic changes in $R_{t}$ over short spans of time, most notably an increase from below one to above 2 at the start of October 2021. 
Overall, we think the EIRR-ww model provides a better estimate of $R_{t}$ than the Huisman method. 
\par
In addition, we calculate the case detection rate normalized by total diagnostic tests to account for changes in the case detection rate due to changes in available diagnostic tests.
The posteriors are visualized in Figure A19. Both normalized and un-normalized versions of the estimated posterior case detection rate changed dramatically during the observation period.
More details are available in Web Section A.2.12.

\section{Discussion}
We have presented a modeling framework for using simplified compartmental models coupled with Bayesian non-parametric priors to estimate the effective reproduction number. 
Using this framework, we created the EIRR-ww model to estimate the effective reproduction number using pathogen genome concentrations collected from wastewater samples. 
We tested the EIRR-ww model by fitting it to data simulated from an agent-based stochastic SEIRR model, and showed it could successfully estimate effective reproduction number dynamics. 
We also used the EIRR-ww model to estimate the effective reproduction number for SARS-CoV-2 in Los Angeles, California, showing it provides plausible estimates of the effective reproduction number when used on real world data. 
\par
Our proposed models ignore the individual time-varying shedding dynamics of pathogen genome concentrations. 
The SEIRR-ww model struggled to estimate the effective reproduction number on simulated data which accounted for the time-varying nature of the data.
In contrast, the EIRR-ww model had smaller bias and credible interval widths and was well calibrated from a frequentist perspective. 
The gap in performance likely stems from the fact that both models use mis-specified observation models and the EIRR-ww model's lack of an $S$ compartment results in more flexibility for its estimates of the effective reproduction number, and thus better performance overall. 
However, the EIRR-ww model was unable to reconstruct the latent population compartments using only the linear relationship between compartment counts and concentrations (Figure A9). 
\par 
Our method clearly outperformed the \citet{huisman2022wastewater} method on both simulated and real data.
We speculate the high levels of noise in both simulated and real data sets resulted in overly jagged estimates from \texttt{EpiEstim}, a problem we demonstrate directly in \citep{goldstein2022incorporating}.
Our method is an appealing alternative when the main goal is effective reproduction number inference. 
\par
When compared to models fit to case data, the EIRR-ww model clearly had larger bias and wider credible intervals, likely due to the high individual variation in genome concentrations seen in empirical studies \citep{hoffmann2021faecal}.
On the other hand, in real world settings where the case detection ratio changes over time, models fit to wastewater data may outperform models fit to case data which do not account for this. 
Using models fit to both cases and wastewater data simultaneously is a promising direction for future work. 
\par
The EIR-cases model had slightly larger bias and wider credible intervals than the SEIR-cases model. However, this comparison was made without considering waning immunity. 
We did not use the SEIR-cases model to estimate $R_{t}$ for Los Angeles, CA, in part because we think it is highly likely the rates of waning immunity changed dramatically when Omicron became the dominant SARS-CoV-2 variant. 
This is a situation for which the EIR framework is well suited. 
Even without wastewater data, we think the framework we describe in this paper is a useful alternative to \texttt{EpiEstim} and related methods when estimating the effective reproduction number.
\par 
We used both the EIRR-ww model as well as three other case based models to estimate the effective reproduction number of SARS-CoV-2 in Los Angeles, CA. 
We emphasize that all methods have their limitations, and none should be taken as ground truth. 
However, the models agreed at many key points in time, including when $R_{t}$ falls below one after the summer 2021 wave, and the general timing of the winter 2021 wave.
This agreement suggests the EIRR-ww model estimates are not unreasonable when fit to real data. 
For a fuller discussion of the points of disagreements between the models, see Web Section A.3.1.
\par 
In this study we focused on replicates, rather than the commonly reported averages of replicates and found that, when using simulated data, models using replicates performed modestly better than models relying on average concentrations \citep{duvallet2022nationwide,wastewaterscan}. 
While the improvements in performance are not large, these improvements are basically free, as the data are already being collected.
We also found that using ten replicates instead of three produced only modestly improved model performance. 
Depending on the cost of producing replicates, increasing the number of replicates sampled may not be worth pursuing. 
\par
For SARS-CoV-2, most shedding occurs in the infectious period, raising the possibility our model could be simplified to exclude the R1 compartment. 
This would simplify our closed form solutions and speed computations, and is a promising modification of the model to explore further. However, including multiple compartments may be necessary for other pathogens with different shedding profiles. We chose not to incorporate covariates that can control for changes in population size and conditions in the sewer system into our model. 
For our particular application, the JWPCP plant is so large, and collects wastewater from so many different smaller plants, that the population size and conditions are plausibly stable across time.
In addition, there remains some controversy over exactly which covariates would be most useful to include \citep{maal2023does}.
Incorporating covariates which adjust for these changes is an important next step.

\section*{Acknowledgments}
This work was in part funded by the UC Irvine Investing to Develop Center-Scale Multidisciplinary Convergence Research Programs Seed Funding Award and by the UC CDPH Modeling Consortium.
DMP was supported by funding from the Bill \& Melinda Gates Foundation (Award: INV-028123) and the NIAID (FAIN: U19AI089672).
SCJ was supported by Water Research Foundation (WRF5093), NSF (CBET 2027306), UC Irvine Clinical Research Acceleration and Facilitation Team (CRAFT), and UC Irvine COVID-19 initiative.
This work utilized the resources of the Research Cyberinfrastructure Center (RCIC) at UC Irvine.
We thank the UC-CDPH modeling consortium, for useful discussion and feedback of intermediate work.
We thank Phil O'Neil and Theodore Kypraios for their workshop teaching materials which inspired our simulation method, and Damon Bayer for visualization code.
We thank the Los Angeles County Sanitation Distribution for providing SARS-CoV-2 viral concentration data.

\bibliographystyle{../../biom}  
\bibliography{../../references}  

\clearpage

\appendix

\setcounter{table}{0}
\setcounter{equation}{0}
\setcounter{section}{0}
\setcounter{figure}{0}

\renewcommand\thefigure{\thesection\-\arabic{figure}}
\renewcommand\thetable{\thesection\-\arabic{table}}

\section{Appendix}
\subsection{Methods}
\subsubsection{Choosing a $\lambda$ prior for SARS-CoV-2 using a stochastic SEIRR} \label{lambda_prior}
The parameter $\lambda$ controls how much pathogen genomic concentrations are attributed to the total number of infectious individuals, versus the total number of recently recovered individuals. 
Previous studies on the timing and magnitude of shedding SARS-CoV-2 RNA have necessarily been concerned with the shedding dynamics of individuals \citep{benefield2020sars,miura2021duration,hoffmann2021faecal}.
However, our model is concerned with the shedding dynamics of populations.
We wanted to create a prior for $\lambda$ which incorporates what we know about shedding dynamics for SARS-CoV-2, but needed a way to translate that information into an appropriate prior for population level shedding dynamics.
Furthermore, our model assumes the relationship between compartment counts and gene concentrations is linear, we wanted to make sure our process for constructing the prior for $\lambda$ incorporated this linearity assumption.
Thus, we used a simulation study using an individual-level engine which could incorporate our prior individual level information combined with linear regression models in order to elicit an appropriate prior for $\lambda$. 
We simulated 1000 epidemics from an agent-based stochastic SEIRR model (the stochastic equivalent to an SEIR model with two R compartments, see Web Section A.2.1), where the times of each individual's transitions between different model states were recorded. 
The population was 1000, $R_{0}$ was set to 2, there were 5 initially infectious agents, and all other agents were susceptible.
For individual $i$ in the $I$ or $R1$ state infected at time $t_{i}$, the concentration of pathogen genomes associated with their shedding at time $l$ was modeled as a random variable $Z_{i}(t_{i}, l)$, where 
\begin{equation*}
    Z_{i}(t_{i}, l) \sim 10^{\text{Normal}(\mu_{i}(l - t_{i}), 1.09)}
\end{equation*}
and the value of $\mu_{i}(l - t_{i})$ was calculated using the consensus shedding load profile of SARS-CoV-2 pathogen RNA generated in \citep{nourbakhsh2022wastewater} by synthesizing previous studies \citep{benefield2020sars, miura2021duration, hoffmann2021faecal}. 
\par 
The value $1.09$ is the average variation in genome concentrations on the log base 10 scale amongst individuals over the course of an infection. 
We calculated this global average variation in concentrations by averaging over the empirical standard deviations of SARS-CoV-2 RNA shed by individuals 6 to 22 days after symptom onset, using the data available from \citep{hoffmann2021faecal}, which uses data collected by \citet{wolfel2020virological}, \citet{han2020viral}, and \citet{lui2020viral}. 
The population level concentration of genomes at time $l$ was the sum of all the individual genome concentrations in the $I$ and $R1$ states at time $l$ divided by the total population size.
\par 
To account for uncertainty in the parameters governing how long individuals spend in the $I$ and $R1$ compartments, for each simulation, $\gamma$, $\nu$ and $\eta$ were chosen from the priors we used when fitting the EIRR model to SARS-CoV-2 data (Table A2).
\par 
Then, for each simulation, we fit a linear model (constrained to positive coefficients using a method described in \citep{glmnet}):
\begin{equation*}
    E[\text{Total genome concentration}] = \beta_{1} \times \text{Prevalence in \textit{I}} + \beta_{2} \times \text{Prevalence in \textit{R1}} 
\end{equation*}
and calculated $\lambda = \beta_{1}/(\beta_{1} + \beta_{2})$. 
Finally, we constructed a logit-normal prior for $\lambda$ which matches the 95\% quantiles of these 1000 $\lambda$ values by minimizing the squared error of the logit-normal prior quantiles and the 95\% quantiles from our 1000 $\lambda$ values using the Nelder-Mead algorithm implemented in the \texttt{optim} function in \texttt{R} \citep{R}. 
Further details of the simulation protocol are available in Web Sections A.2.1, A.2.2, and A.2.3.

\subsubsection{Details of the Huisman Method}
We compare the EIRR-ww model to the \citet{huisman2022wastewater} method.
This method is a variation on the well known \texttt{EpiEstim} method \citep{cori_new_2013}. 
Pathogen genome concentrations are modeled as a function of incidence (newly infected individuals) counts via a convolution equation:
\begin{equation}
    C_{i} = M \sum_{j} w_{i-j}I_{j}.
\end{equation}
Here $C_{i}$ is the concentration at time $i$, $I_{j}$ is the number of new infections in the period $(j-1,j]$, and $w_{i-j}$ is a weight derived from discretizing the assumed individual shedding load profile describing how many pathogen genomes an infected individual sheds over time. 
$M$ is a constant value translating counts of individuals to counts of pathogens, the Huisman method assumes $M$ is the lowest observed concentration in the data set.
A time series of incidence is constructed via a deconvolution algorithm, and then used as the inputs into \texttt{EpiEstim}.
\texttt{EpiEstim} is a method inspired by branching process approximations of the spread of infectious disease where infectious individuals generate new infectious individuals in a Crump-Mode-Jager process \citep{Fraser2007,cori_new_2013,pakkanen2023unifying}. 
The core concept is to model current incidence as a function of previous incidence and the effective reproduction number through the so-called Renewal Equation:
\begin{equation}
    \text{E}(I_{t} \mid \mathbf{I}_{1:t},R_{t}) = R_{t}\sum_{u=1}^{t-1}I_{u}g_{t-u}.
\end{equation}
Here $g_{t-u}$ are values from the discretized generation time distribution, the distribution of the time between one person becoming infected, and subsequently infecting someone else. 
\texttt{EpiEstim} models incidence conditioned on previous incidence and the effective reproduction number as a Poisson random variable, and holds $R_{t}$ constant for a window of time, creating a smooth estimate by repeatedly re-estimating $R_{t}$ for all such windows in the time series. 
\citet{huisman2022wastewater} repeats this pipeline multiple times via a bootstrap method to generate uncertain estimates of the effective reproduction number.

\subsubsection{Priors for models with the $S$ compartment}\label{seirr_priors}
We will assume the basic reproduction number is constant in a time interval $(k_{i}, k_{i+1}]$, defining it as $R_{0,k_{i}} = \frac{\beta_{k_{i}}}{\nu}$. 
Let $M$ be the total number of time intervals of interest.
Let $\mathbf{R_{0}} = (R_{0,0}, R_{0,k_{1}}, \dots, R_{0, k_{M}})$, be the vector of basic reproduction numbers. 
We use a random walk prior so that
\begin{align*}
R_{0,0} &\sim \text{Log-Normal}(\mu_{0,0}, \sigma_{0,0}), \\
\sigma &\sim \text{Log-Normal}(\mu_{rw}, \sigma_{rw}), \\
\log{(R_{0,k_{i+1}})}|\log{(R_{0,k_{i}})},\sigma  &\sim \text{Normal}(\log{(R_{0,k_{i}})}, \sigma_{rw}). \\
\end{align*}
For this study, we assume the basic reproduction number changes on a weekly basis, but it could change according to other time scales.
Our model also requires initial conditions in order to solve the system of ODEs. 
Let $N$ be the population size (assumed to be known). 
Let $P$ be the population not in the $R2$ compartment at the time the model is fit. 
For simulations, this value is also known, in real world settings, we assume $N=P$, that is, the difference between the two is negligible in large populations. 
In the case of the SEIRR-ww model, the initial conditions are calculated as:
\begin{align*}
S(0) &= P * S\_SEIR1, \\
I(0) &= (P - S(0)) * I\_EIR1, \\
R1(0) &= (P - S(0) - I(0)) * R1\_ER1, \\
E(0) &= (P - S(0) - I(0) - R1(0)), \\
R2(0) &= 1,
\end{align*}
where we define $S\_SEIR1$ as the proportion of the population in the $S$ compartment at time $0$, $I\_EIR1$ as the proportion of those in the $E$, $I$, or $R1$ compartments in the $I$ compartment and $R1\_ER1$ as the proportion of those in the $E$ or $R1$ compartments in the $R1$ compartment. 
We use logit-normal priors for the proportions. 
We use a similar technique when using the SEIR-cases model, with one less parameter as there is only one $R$ compartment.

\subsubsection{Closed form solutions of the EIR/EIRR models}
The systems of ordinary differential equations for the EIR/EIRR models are linear, and thus can be solved in closed form. 
We used \texttt{Mathematica} Version (13.1) to calculate the closed form solution. 

\subsubsection{Closed form solution for the EIRR Model}\label{sec:closed_form}
Define the EIRR model as: 
\begin{align*}
    \frac{dE}{dt} &= \alpha_{t}*I - \gamma E\\
    \frac{dI}{dt} &= \gamma E - \nu I \\
    \frac{dR1}{dt} &= \nu I - \eta R1\\
    \frac{dR2}{dt} &= \eta R1
\end{align*}

Let $V$ be 
\begin{equation*}
    V = \begin{bmatrix}
    -\gamma & \alpha_{t} & 0 & 0\\
    \gamma &  -\nu & 0 & 0\\
    0 & \nu & -\eta & 0 \\
    0 & 0 & \eta & 0
    \end{bmatrix}.
\end{equation*}
The matrix exponential of $V$ for a fixed $\alpha_{t}$ is reported on the next page.
For initial conditions $M(t_{0})$, the solution to the system of ODEs is 
\begin{align*}
    e^{V(t-t_{0})}M(t_{0}).
\end{align*}
\newpage 
  \setlength{\arraycolsep}{1pt}
  \renewcommand{\arraystretch}{0.4}
{\tiny 
\noindent $a = 4 \alpha  \gamma +\gamma ^2-2 \gamma  \nu +\nu ^2$\\
$b = 3 \eta +3 \gamma +3 \nu$\\
$c = -2 \alpha 
   \gamma +2 \eta  \gamma +2 \eta  \nu +2 \gamma  \nu $
\\
$A =  \frac{\left(\frac{1}{2} (\alpha  \gamma -\gamma  (\eta -\gamma )) \left(-\sqrt{a}-\gamma +\nu \right)+\gamma  (\alpha  (\eta -\nu
   )-\alpha  \gamma )\right) e^{\frac{1}{2} t \left(-\sqrt{a}-\gamma -\nu \right)}}{\frac{1}{4} \left(-\sqrt{a}-\gamma -\nu \right)^2 (b )+\frac{1}{2} \left(-\sqrt{a}-\gamma -\nu \right) (c )-\alpha  \eta  \gamma +\frac{1}{2} \left(-\sqrt{a}-\gamma -\nu \right)^3+\eta 
   \gamma  \nu }+\frac{\left(\frac{1}{2} (\alpha  \gamma -\gamma  (\eta -\gamma )) \left(\sqrt{a}-\gamma +\nu \right)+\gamma  (\alpha 
   (\eta -\nu )-\alpha  \gamma )\right) e^{\frac{1}{2} t \left(\sqrt{a}-\gamma -\nu \right)}}{\frac{1}{4} \left(\sqrt{a}-\gamma -\nu \right)^2 (b )+\frac{1}{2} \left(\sqrt{a}-\gamma -\nu \right) (c )-\alpha  \eta  \gamma +\frac{1}{2} \left(\sqrt{a}-\gamma -\nu
   \right)^3+\eta  \gamma  \nu }$
\\
$B = \frac{\left(\frac{1}{2} \left(-\sqrt{a}+\gamma -\nu \right) (\alpha  (\eta -\nu )-\alpha  \gamma
   )+\alpha  (\alpha  \gamma -\gamma  (\eta -\gamma ))\right) e^{\frac{1}{2} t \left(-\sqrt{a}-\gamma -\nu \right)}}{\frac{1}{4}
   \left(-\sqrt{a}-\gamma -\nu \right)^2 (b )+\frac{1}{2} \left(-\sqrt{a}-\gamma -\nu \right) (c )-\alpha  \eta  \gamma +\frac{1}{2} \left(-\sqrt{a}-\gamma -\nu \right)^3+\eta  \gamma  \nu }+\frac{\left(\frac{1}{2} \left(\sqrt{a}+\gamma -\nu \right) (\alpha  (\eta -\nu
   )-\alpha  \gamma )+\alpha  (\alpha  \gamma -\gamma  (\eta -\gamma ))\right) e^{\frac{1}{2} t \left(\sqrt{a}-\gamma -\nu
   \right)}}{\frac{1}{4} \left(\sqrt{a}-\gamma -\nu \right)^2 (b )+\frac{1}{2} \left(\sqrt{a}-\gamma -\nu \right) (c )-\alpha  \eta  \gamma +\frac{1}{2} \left(\sqrt{a}-\gamma -\nu \right)^3+\eta  \gamma  \nu }$
   \\
$C = \frac{\left(\frac{1}{2} \left(-\sqrt{a}-\gamma +\nu \right) (\gamma  (\eta -\gamma )-\gamma  \nu )+\gamma  (\alpha  \gamma -\nu  (\eta
   -\nu ))\right) e^{\frac{1}{2} t \left(-\sqrt{a}-\gamma -\nu \right)}}{\frac{1}{4} \left(-\sqrt{a}-\gamma -\nu \right)^2 (b )+\frac{1}{2} \left(-\sqrt{a}-\gamma -\nu \right) (c )-\alpha  \eta  \gamma +\frac{1}{2} \left(-\sqrt{a}-\gamma -\nu \right)^3+\eta  \gamma  \nu
   }+\frac{\left(\frac{1}{2} \left(\sqrt{a}-\gamma +\nu \right) (\gamma  (\eta -\gamma )-\gamma  \nu )+\gamma  (\alpha  \gamma -\nu 
   (\eta -\nu ))\right) e^{\frac{1}{2} t \left(\sqrt{a}-\gamma -\nu \right)}}{\frac{1}{4} \left(\sqrt{a}-\gamma -\nu \right)^2 (b )+\frac{1}{2} \left(\sqrt{a}-\gamma -\nu \right) (c )-\alpha  \eta  \gamma +\frac{1}{2} \left(\sqrt{a}-\gamma -\nu \right)^3+\eta 
   \gamma  \nu }$ \\
$D = \frac{\left(\frac{1}{2} \left(-\sqrt{a}+\gamma -\nu \right) (\alpha  \gamma -\nu  (\eta -\nu ))+\alpha  (\gamma  (\eta
   -\gamma )-\gamma  \nu )\right) e^{\frac{1}{2} t \left(-\sqrt{a}-\gamma -\nu \right)}}{\frac{1}{4} \left(-\sqrt{a}-\gamma -\nu \right)^2 (b )+\frac{1}{2} \left(-\sqrt{a}-\gamma -\nu \right) (c )-\alpha  \eta  \gamma +\frac{1}{2} \left(-\sqrt{a}-\gamma -\nu
   \right)^3+\eta  \gamma  \nu }+\frac{\left(\frac{1}{2} \left(\sqrt{a}+\gamma -\nu \right) (\alpha  \gamma -\nu  (\eta -\nu ))+\alpha 
   (\gamma  (\eta -\gamma )-\gamma  \nu )\right) e^{\frac{1}{2} t \left(\sqrt{a}-\gamma -\nu \right)}}{\frac{1}{4} \left(\sqrt{a}-\gamma -\nu \right)^2 (b )+\frac{1}{2} \left(\sqrt{a}-\gamma -\nu
   \right) (c )-\alpha  \eta  \gamma +\frac{1}{2} \left(\sqrt{a}-\gamma -\nu
   \right)^3+\eta  \gamma  \nu }$
   \\
$E = \frac{\gamma  \nu  e^{-\eta  t}}{-\alpha  \gamma +\eta ^2-\eta  \gamma -\eta  \nu +\gamma  \nu }+\frac{\left(-\gamma  \nu  \sqrt{a}+\gamma ^2 (-\nu )-\gamma  \nu ^2\right) e^{\frac{1}{2} t \left(-\sqrt{a}-\gamma -\nu \right)}}{2 \left(\frac{1}{4} \left(-\sqrt{4
   \alpha  \gamma +\gamma ^2-2 \gamma  \nu +\nu ^2}-\gamma -\nu \right)^2 (b )+\frac{1}{2} \left(-\sqrt{a}-\gamma
   -\nu \right) (c )-\alpha  \eta  \gamma +\frac{1}{2} \left(-\sqrt{a}-\gamma
   -\nu \right)^3+\eta  \gamma  \nu \right)}+ \\
   \frac{\left(\gamma  \nu  \sqrt{a}+\gamma ^2 (-\nu )-\gamma  \nu ^2\right) e^{\frac{1}{2} t
   \left(\sqrt{a}-\gamma -\nu \right)}}{2 \left(\frac{1}{4} \left(\sqrt{a}-\gamma -\nu
   \right)^2 (b )+\frac{1}{2} \left(\sqrt{a}-\gamma -\nu \right) (c )-\alpha  \eta  \gamma +\frac{1}{2} \left(\sqrt{a}-\gamma -\nu \right)^3+\eta  \gamma  \nu \right)}$
   \\
$F = -\frac{e^{-\eta  t}
   \left(\eta ^2 \nu -\eta  \gamma  \nu \right)}{\eta  \left(-\alpha  \gamma +\eta ^2-\eta  \gamma -\eta  \nu +\gamma  \nu \right)}+\frac{\left(\nu ^2 \sqrt{a}+2 \alpha  \gamma  \nu -\gamma  \nu ^2+\nu ^3\right) e^{\frac{1}{2} t \left(-\sqrt{a}-\gamma -\nu \right)}}{2
   \left(\frac{1}{4} \left(-\sqrt{a}-\gamma -\nu \right)^2 (b )+\frac{1}{2} \left(-\sqrt{a}-\gamma -\nu \right) (c )-\alpha  \eta  \gamma +\frac{1}{2} \left(-\sqrt{a}-\gamma -\nu \right)^3+\eta  \gamma  \nu \right)}+\frac{\left(-\nu ^2 \sqrt{a}+2 \alpha  \gamma  \nu -\gamma 
   \nu ^2+\nu ^3\right) e^{\frac{1}{2} t \left(\sqrt{a}-\gamma -\nu \right)}}{2 \left(\frac{1}{4} \left(\sqrt{a}-\gamma -\nu \right)^2 (b )+\frac{1}{2} \left(\sqrt{a}-\gamma -\nu \right) (c )-\alpha  \eta  \gamma +\frac{1}{2} \left(\sqrt{a}-\gamma -\nu \right)^3+\eta 
   \gamma  \nu \right)}$
   \\
$G = \frac{e^{-\eta  t} \left(-\sqrt{a}+2 \eta -\gamma -\nu \right) \left(\sqrt{a}+2 \eta -\gamma -\nu \right)}{4 \left(-\alpha  \gamma +\eta ^2-\eta  \gamma -\eta  \nu +\gamma  \nu \right)}$
\\
$H = -\frac{\gamma  \nu  e^{-\eta  t}}{-\alpha  \gamma +\eta ^2-\eta  \gamma -\eta  \nu +\gamma  \nu }+\frac{\eta  \gamma  \nu  e^{\frac{1}{2} t \left(-\sqrt{a}-\gamma -\nu \right)}}{\frac{1}{4} \left(-\sqrt{a}-\gamma -\nu \right)^2 (b )+\frac{1}{2}
   \left(-\sqrt{a}-\gamma -\nu \right) (c )-\alpha  \eta  \gamma +\frac{1}{2}
   \left(-\sqrt{a}-\gamma -\nu \right)^3+\eta  \gamma  \nu }+\frac{\eta  \gamma  \nu  e^{\frac{1}{2} t \left(\sqrt{a}-\gamma -\nu \right)}}{\frac{1}{4} \left(\sqrt{a}-\gamma -\nu \right)^2 (b
   )+\frac{1}{2} \left(\sqrt{a}-\gamma -\nu \right) (c )-\alpha  \eta  \gamma
   +\frac{1}{2} \left(\sqrt{a}-\gamma -\nu \right)^3+\eta  \gamma  \nu }+\frac{\eta  \gamma  \nu }{\eta  \gamma  \nu -\alpha  \eta 
   \gamma }$
   \\
$I = \frac{\nu  (\eta -\gamma ) e^{-\eta  t}}{-\alpha  \gamma +\eta ^2-\eta  \gamma -\eta  \nu +\gamma  \nu }-\frac{\eta  \nu  \left(\sqrt{a}-\gamma +\nu \right) e^{\frac{1}{2} t \left(-\sqrt{a}-\gamma -\nu \right)}}{2 \left(\frac{1}{4} \left(-\sqrt{a}-\gamma -\nu \right)^2 (b )+\frac{1}{2} \left(-\sqrt{a}-\gamma -\nu
   \right) (c )-\alpha  \eta  \gamma +\frac{1}{2} \left(-\sqrt{a}-\gamma -\nu \right)^3+\eta  \gamma  \nu \right)}- \\
   \frac{\eta  \nu  \left(-\sqrt{a}-\gamma +\nu \right) e^{\frac{1}{2} t \left(\sqrt{a}-\gamma -\nu \right)}}{2 \left(\frac{1}{4} \left(\sqrt{a}-\gamma -\nu \right)^2 (3 \eta +3
   \gamma +3 \nu )+\frac{1}{2} \left(\sqrt{a}-\gamma -\nu \right) (c )-\alpha
    \eta  \gamma +\frac{1}{2} \left(\sqrt{a}-\gamma -\nu \right)^3+\eta  \gamma  \nu \right)}+\frac{\eta  \gamma  \nu }{\eta  \gamma  \nu
   -\alpha  \eta  \gamma }$
   \\
$J = 1-\frac{e^{-\eta  t} \left(-\sqrt{a}+2 \eta -\gamma -\nu \right) \left(\sqrt{a}+2 \eta -\gamma -\nu \right)}{4 \left(-\alpha  \gamma +\eta ^2-\eta  \gamma -\eta  \nu +\gamma  \nu \right)}$
\\
$K = -\frac{\eta  \left(\sqrt{a}-\gamma -\nu \right) \left(\sqrt{a}+\gamma +\nu \right)}{4 (\eta  \gamma  \nu -\alpha  \eta  \gamma )}$
\\
$e^{V} = \left(
\begin{array}{cccc}
 A & B & 0 & 0 \\
 C & D & 0 & 0 \\
 E & F & G & 0 \\
 H & I & J & K \\
\end{array}
\right)$}
\newpage 

\subsubsection{Choosing parameters for the Huisman Method when fitting to simulated data}
To choose a shedding load profile, we first re-fit a spline to the points from the \citet{nourbakhsh2022wastewater} profile raised to the tenth power, with an additional value of 0 at time 0. 
We then generated predictions from the spline from 0 to 29 evenly spaced by 0.1, and used these as true values from the shedding load profile.
We then used the Nelder-Mead algorithm to search for shape and scale parameters of the gamma distribution which minimized the squared loss of the proposed grid point values versus our generated true values, and used these parameters as the shape and scale of the shedding load profile for the Huisman model. 
In an SEIR model, the intrinsic generation time distribution is the sum of the latent and infectious periods \citep{Svensson2007,champredon2015,champredon2018}, which is a hypo-exponential distribution. 
\texttt{EpiEstim} is normally used assuming the generation time distribution is a gamma distribution. We used a gamma distribution with mean and standard deviation equal to the true intrinsic generation time distribution. 

\subsubsection{Branching process inspired models}
For additional comparisons of our estimates of SARS-CoV-2 in Los Angeles, CA, we use two branching process inspired models that, unlike compartmental models, only model latent incidence.
The \citet{huisman2022wastewater} method uses one example of this class of methods, but there are many others.
The method relies on the so-called renewal equation which calculates current incidence as a product of a weighted sum of previous incidence and the effective reproduction number. 
Let $I_{t}$ be the incidence at time $t$, $R_{t}$ be the effective reproduction number at time $t$, and $g(t)$ be the probability density function of the generation time distribution (the time between an individual becoming infected and infecting another individual; under the compartmental model framework this is usually taken to be equivalent to the sum of the latent period and the infectious period \citep{Svensson2007, champredon2015, champredon2018}). Then the classic renewal equation is:
\begin{equation*}
    E[I_{t}|I_{1}, \dots, I_{t-1}] = R_{t}\sum_{s=1}^{t-1}I_{s}g(t-s).
\end{equation*}
The \texttt{epidemia} package can be used to create different branching process inspired models to estimate the effective reproduction number using different observation models and models for latent incidence \citep{epidemia}.
For the model we used in this study, we modeled observed cases using a negative binomial distribution, modeled the effective reproduction number as a Gaussian random walk, and modeled unobserved incidence as an auto-regressive normal random variable with variance equal to the mean multiplied by an over-dispersion parameter.
The explicit model is listed below:
 \begin{align*}
\tau &\sim \text{exp}(\lambda)  \text{--Hyperprior for unobserved incidence,}\\
I_{\nu} &\sim \text{exp}(\tau) \text{--Prior on unobserved incidence $\nu$ days before observation,}\\
I_{\nu+1}, \dots, I_{0} &= I_{\nu}  \text{--Unobserved incidence,}\\
    \sigma & \sim \text{Truncated-Normal}(0, 0.1^{2})\text{--Prior on variance of random walk} \\
   \log{R_{0}} &\sim \text{Normal}(\log{2}, 0.2^{2})\text{--Prior on $R_{0}$,} \\
    \log{R_{t}}|\log{R_{t-1}} &\sim \text{Normal}(\log{R_{t-1}}, \sigma) \text{--Random walk prior on $R_{t}$,}\\
\psi &\sim \text{Normal}(10,2) \text{--Prior on variance parameter for incidence,} \\
I_{t}|I_{\nu}, \dots, I_{t-1} &\sim \text{Normal}(R_{t}\sum_{s<t}I_{s}g_{t-s}, \psi) \text{--Model for incidence,} \\
    \alpha &\sim \text{Normal}(0.13, 0.7^2)
\text{--Prior on case detection rate,} \\
y_{t} &= \alpha_{t}\sum_{s<t}I_{s}\pi_{t-s} \text{--Mean of observed data model,}\\
\phi & \sim P(\phi) \text{--Prior on dispersion parameter for observed data,} \\
Y_{t} &\sim \text{Neg-Binom}(y_{t}, \phi) \text{--Observed data model.}\\
\end{align*}
Here $\pi_{t}$ are the values of the probability density function for the delay distribution, the time between an individual being infected and being observed.
We also used the Rt-estim-gamma model, which is similar to the model above, but uses total diagnostic tests as a model covariate in the observation model. 
This allows the rate of detection to change over time as a function of available tests, avoiding the situation, for example, where an increase in cases due to test availability is mistaken for an increase in cases due to increased incidence.
Full details are available in \citep{goldstein2022incorporating}.

\subsection{Results}
\subsubsection{Simulating an agent-based stochastic SEIRR model}\label{sec:sim_method}
\begin{figure}[H]
    \centering
    \includegraphics{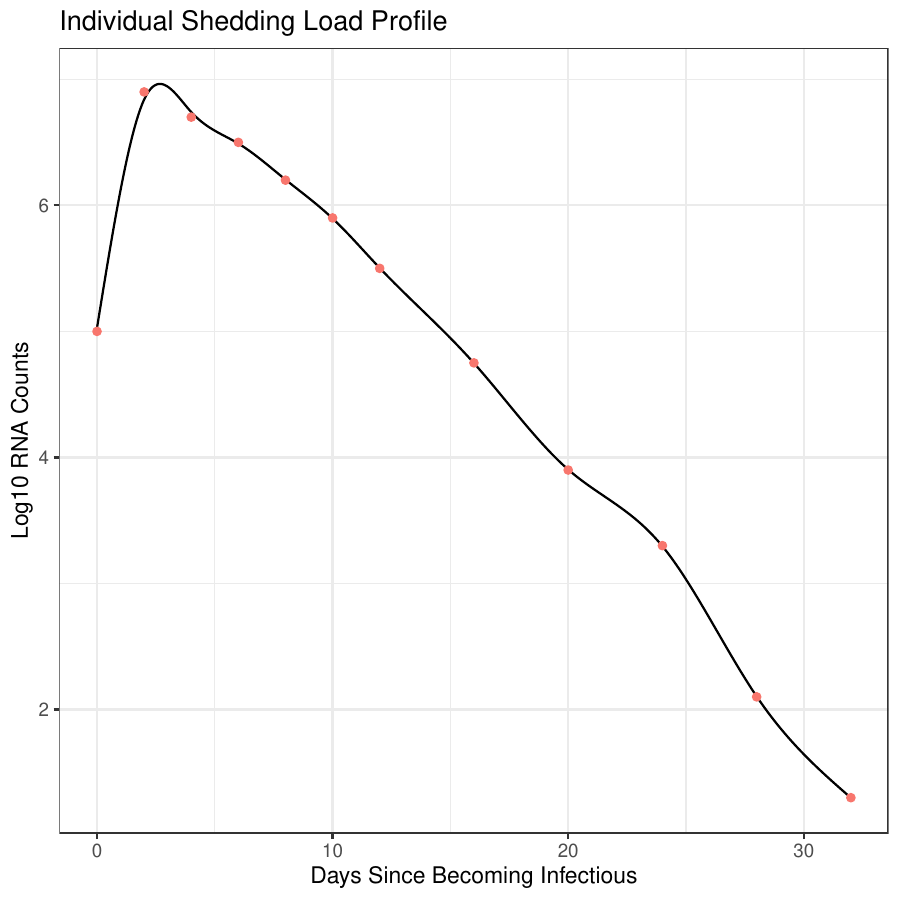}
    \caption{Thin plate spline fit to manually recorded values from the consensus shedding load profile developed by Nourbakhsh et al. (2022). 
    Red dots are the manually recorded values. Black line is the spline fit.}
    \label{fig:spline_fit}
\end{figure}

An agent-based stochastic SEIRR model is an N-dimensional continuous time Markov chain, where N is the population. 
When represented as a vector $G(t)$, each entry of $G(t)$ records the state of one of the $N$ individuals, i.e. if the $ith$ entry of $G(t)_{i}$, $G(t)_{i}$ is $S$, then the $ith$ individual is susceptible at time $t$. 
Let $I(t)$ be the number of infectious individuals at time $t$
It can be defined in terms of its transition rates from state $G$ to state $G'$ so that
\begin{equation*}
\lambda_{GG'} =
    \begin{cases}
        \beta_{t}/N \times I(t) \text{ if $G_{j} = S$ and $G'_{j} = E$} \\
        \gamma \text{ if $G_{j} = E$ and $G'_{j} = I$} \\
        \nu \text{ if $G_{j} = I$ and $G'_{j} = R1$} \\
        \eta \text{ if $G_{j} = R1$ and $G'_{j} = R2$} \\
        0 \text{ otherwise}
    \end{cases}
\end{equation*}
The well known Gillespie algorithm popularized in \citep{gillespie1977exact} can be used to simulate from this model, but it is quite slow. 
We employ a variation of the Gillespie algorithm in order to simulate individuals and their individual state transition times. 
In essence, as an individual enters the simulation (via infection) all future transition times for that individual are simulated at once. 
Then, the next event is simply the most recent transition time amongst all individuals still in the simulation.
In psuedocode, the basic algorithm is
\begin{algorithm}[H]
\caption{Individual Gillespie Algorithm}\label{alg:cap}
\begin{algorithmic}
\State $i \gets \text{initial I}$ 
\State $e \gets \text{initial E}$ 
\State $r1 \gets 0$ 
\State $s \gets N - i - e$
\State $t_{i} \gets \text{Initial infectious times}$
\State $t_{r1} \gets \text{Initial recover times}$
\State $t_{r2} \gets \text{Initial stop shedding times}$
\State $t \gets \max{t_{i}}$
\While{$i > 0 \hspace{1pt}| \hspace{1pt} r1 > 0$}
\If{$s >0 \hspace{1pt} \& \hspace{1pt} i>0$}
    \State $n_{e} \gets t + \text{Exp}(\beta \times s \times i)$ 
\ElsIf{$s = 0 \hspace{1pt} | \hspace{1pt} i = 0$}
    \State $n_{e} \gets \inf$
\EndIf
\State $n_{i} \gets \min{t_{i}}$
\State $n_{r1} \gets \min{t_{r1}}$
\State $n_{r2} \gets  \min{t_{r2}}$
\State $n_{t} \gets \min{n_{e}, n_{i}, n_{r1}, n_{r2}}$
\State $t \gets n_{t}$
\If{$n_{t} = n_{e}$}
\State $s \gets s - 1$
\State $e \gets e + 1$
\State $x \sim \text{Exp}(\gamma)$
\State $y \sim \text{Exp}(\nu)$
\State $z \sim \text{Exp}(\eta)$
\State $t_{i} \gets \text{append}(t_{i}, t + x)$
\State $t_{r1} \gets \text{append}(t_{r1}, t + x + y)$
\State $t_{r2} \gets \text{append}(t_{r2}, t + x + y + z)$
\ElsIf{$n_{t} = n_{i}$}
\State $e \gets e - 1$
\State $i \gets i + 1$
\State $t_{i} \gets \text{remove}(t_{i}, n_{i})$
\ElsIf{$n_{t} = n_{r1}$}
\State $i \gets i - 1$
\State $r1 \gets r1 + 1$
\State $t_{r1} \gets \text{remove}(t_{r1}, n_{r1})$
\ElsIf{$n_{t} = n_{r2}$}
\State $r1 \gets r1 - 1$
\State $t_{r2} \gets \text{remove}(t_{r2}, n_{r2})$
\EndIf
\EndWhile
\end{algorithmic}
\end{algorithm}
We omit the various pieces of code which ensure that all of the individual transition times are recorded and associated with the correct individuals. 
It is easy to adapt this algorithm to allow for a changing $\beta$ at known change point times. 
We simply add an additional check, where the next event can be any of the nearest future transition times, or the nearest future change point time.
We use this adapted version when we simulate data to test the EIRR-ww model.
In order to calculate individual genome concentrations, we use the consensus shedding profile for SARS-CoV-2 RNA created by \citet{nourbakhsh2022wastewater}. 
We manually recorded the values displayed in the figure, and then used a thin plate regression spline (using the \texttt{R} package \texttt{fields} \citep{fields}) to create a continuous curve. 
The spline fit and manually recorded values are displayed in Figure \ref{fig:spline_fit}. 

To calculate the mean log genome concentration shed by an individual at time time $l$, we first checked to see if they were in the $I$ or $R1$ compartments at time $l$, then predicted mean log genome concentration using the fitted spline and the difference between $l$ and the time they became infectious. 

\subsubsection{Comparing Simulation Engines}
We compare our variation on the Gillespie algorithm to both a traditional Agent based model (Agent) and a compartmental model (Compartment) simulated using the traditional Gillespie algorithm. 
We set the population to be 100, set $R_{0}$ to 1.5 and initialized with 5 infectious individuals and all other individuals in the susceptible population. 
We simulated 10000 simulations from each of the three engines, and plotted the mean and quantiles of the counts in each of the compartments in the first 100 days. 
For the agent and variation models, we also plotted the log concentrations on each of the first 100 days. 
We see no evidence the three engines are not equivalent. 
\begin{figure}[H]
    \centering
    \includegraphics[width = \textwidth]{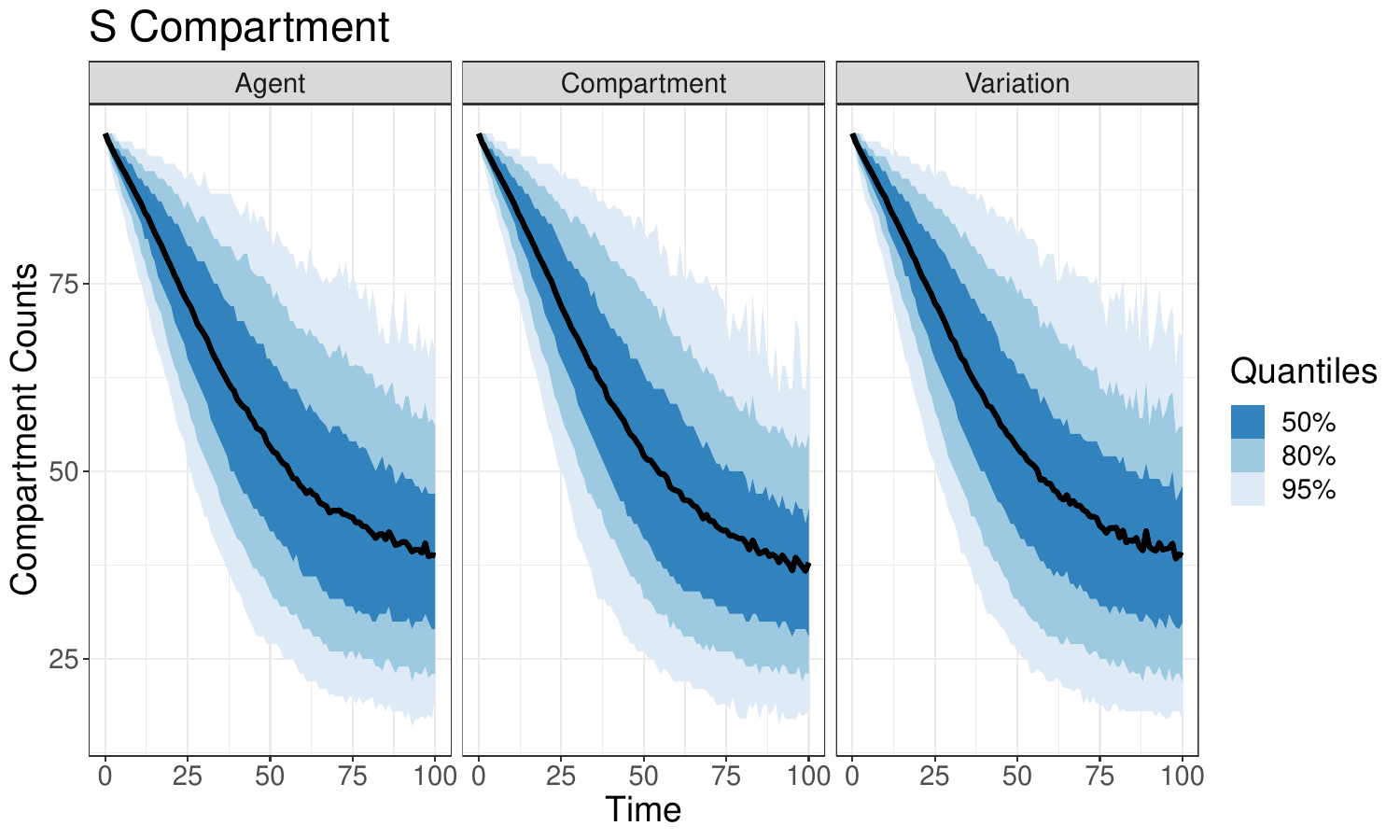}
    \caption{Counts in the S compartment across three simulation engines. Blue bars are quantiles, black lines are means.}
    \label{fig:S_mc_comp}
\end{figure}
\begin{figure}[H]
    \centering
    \includegraphics[width = \textwidth]{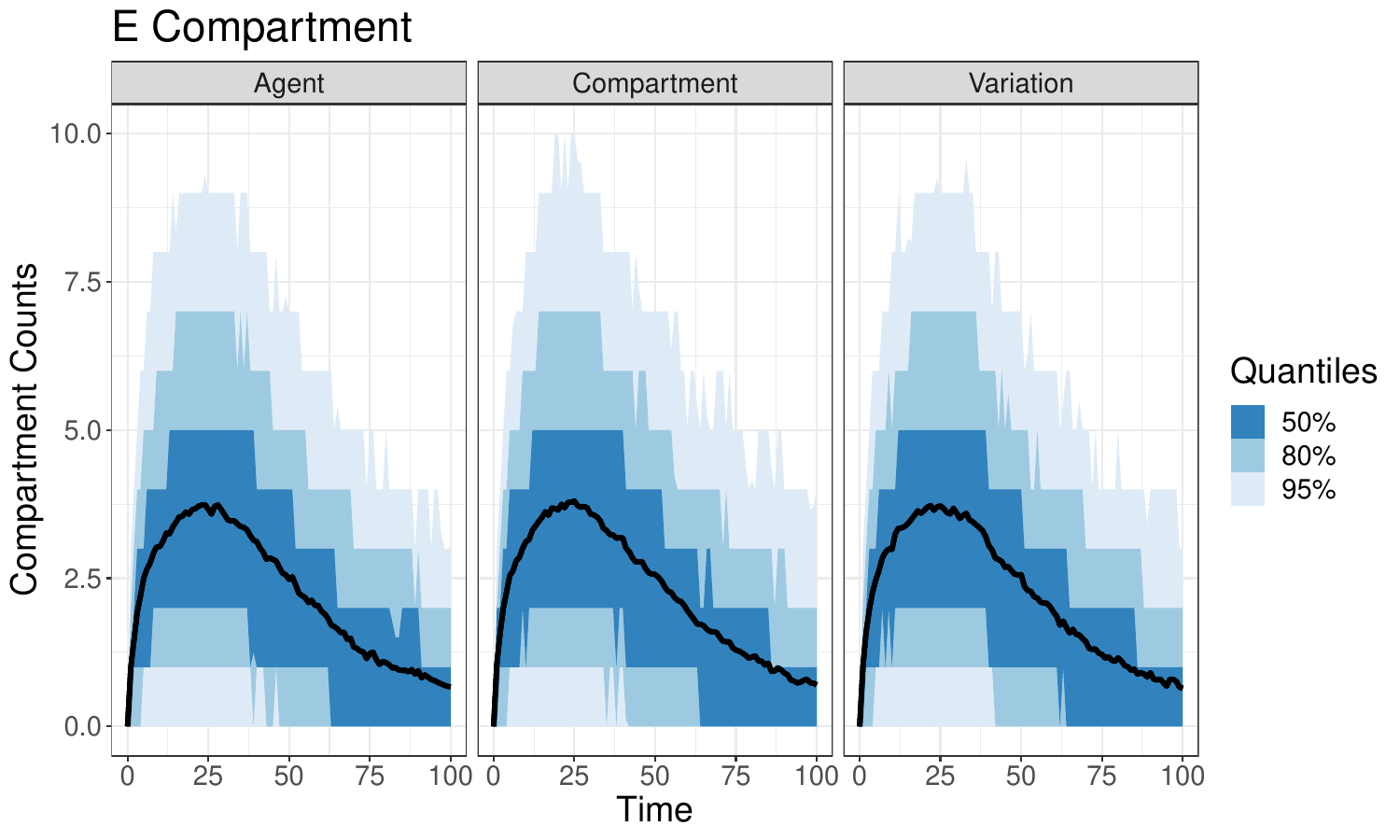}
    \caption{Counts in the E compartment across three simulation engines. Blue bars are quantiles, black lines are means.}
    \label{fig:E_mc_comp}
\end{figure}
\begin{figure}[H]
    \centering
    \includegraphics[width = \textwidth]{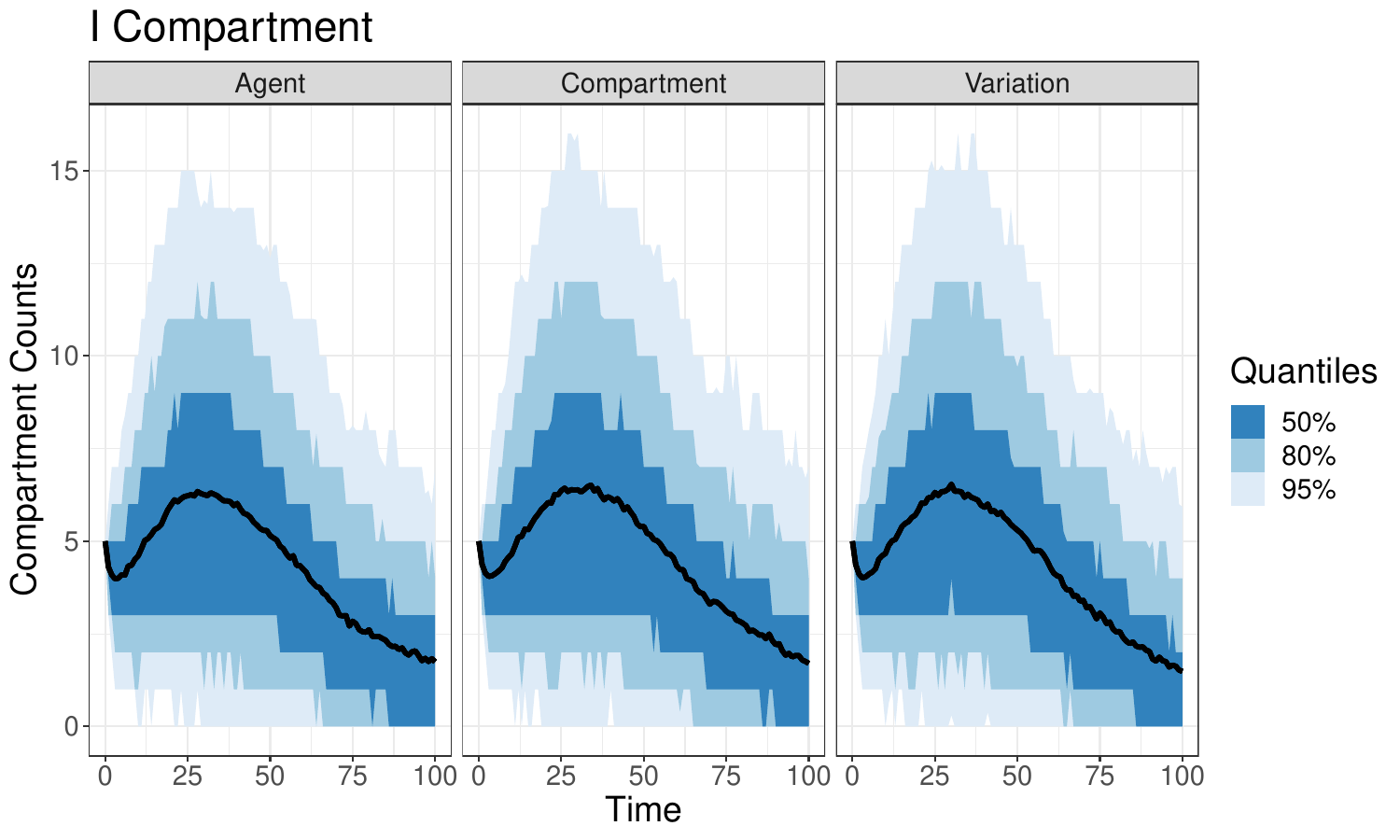}
    \caption{Counts in the I compartment across three simulation engines. Blue bars are quantiles, black lines are means.}
    \label{fig:I_mc_comp}
\end{figure}
\begin{figure}[H]
    \centering
    \includegraphics[width = \textwidth]{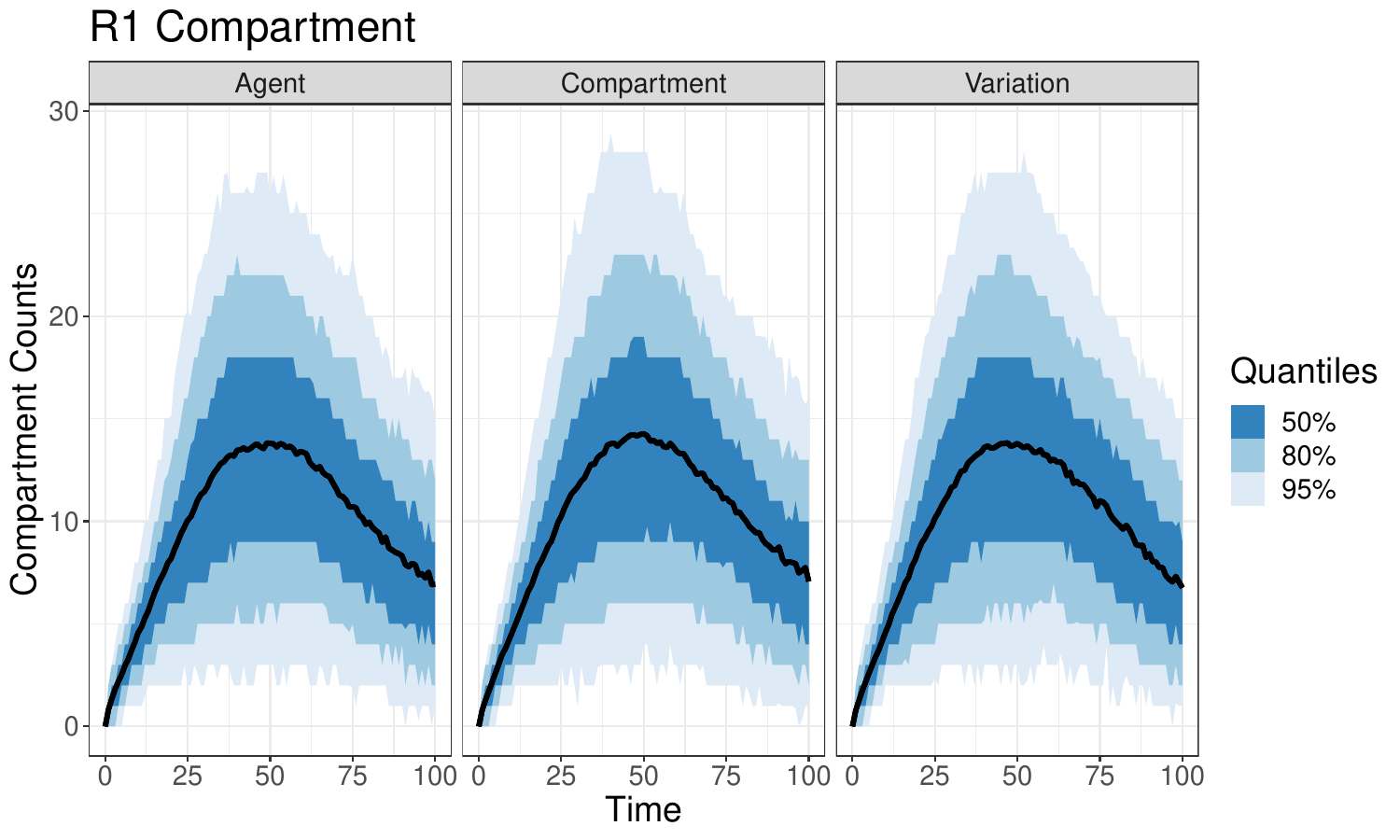}
    \caption{Counts in the R1 compartment across three simulation engines. Blue bars are quantiles, black lines are means.}
    \label{fig:R1_mc_comp}
\end{figure}
\begin{figure}[H]
    \centering
    \includegraphics[width = \textwidth]{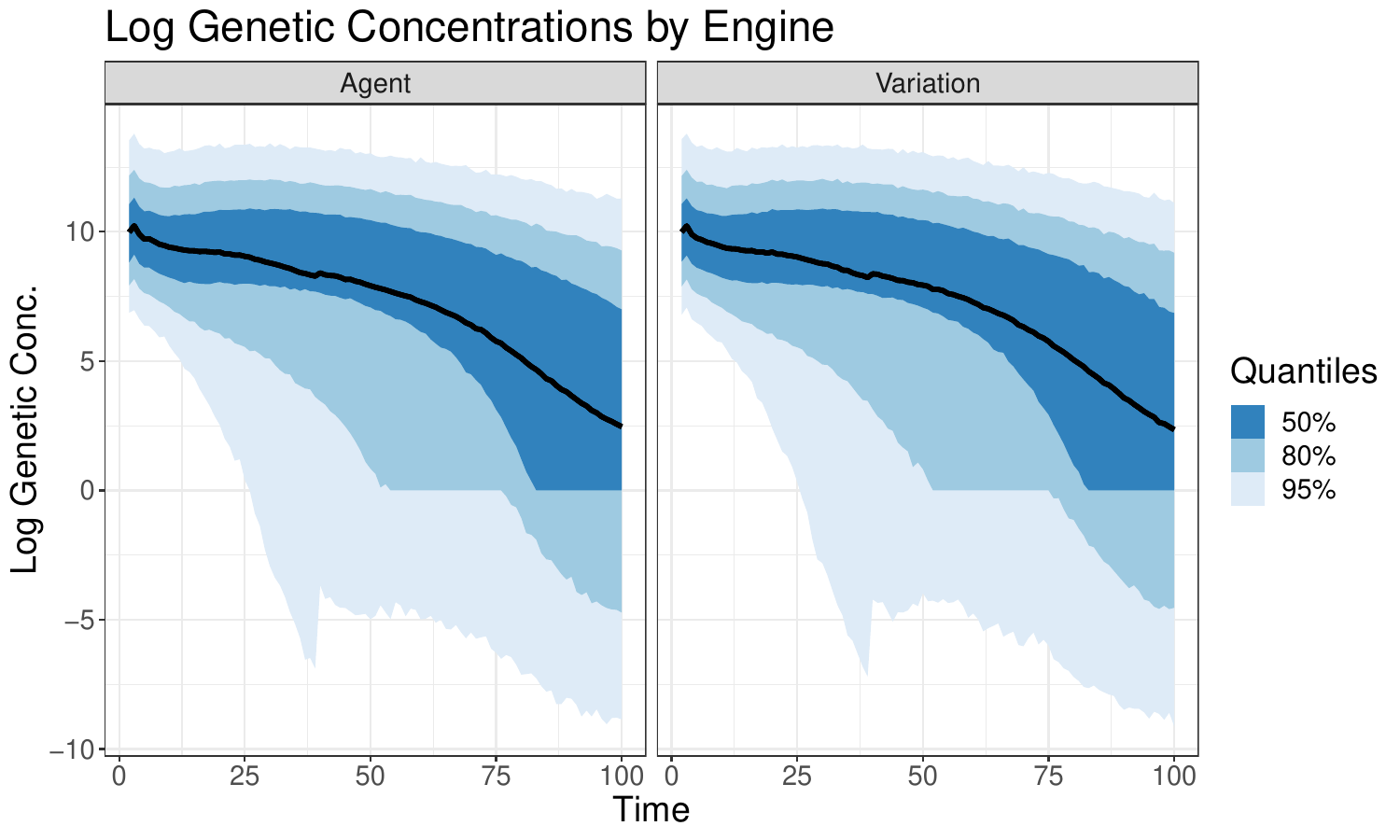}
    \caption{Simulated log concentrations across two simulation engines. Blue bars are quantiles, black lines are means.}
    \label{fig:conc_mc_comp}
\end{figure}

\subsubsection{Simulation parameters}\label{sec:sim_params}
For choosing the prior for $\lambda$, the population was 1000, $R_{0}$ was set to 2, and the parameters for $\gamma$, $\nu$ and $\eta$ were drawn from the priors in Table \ref{tab:ww_sim_priors}. 
Five individuals were initially in the $I$ compartment, the other individuals were in the $S$ compartment.
\par 
The parameters used for the simulation producing observed data used in the simulation studies described in the main text are displayed below. 
The population size was $100000$, the model was initialized with $200$ individuals in the $E$ compartment and $200$ in the $I$ compartment, all others were in the $S$ compartment.
The basic reproduction number was set to 1.75 for five weeks, $0.7$ for four weeks, and then increased from 0.7 to 0.9 for two weeks before following the trajectory chosen for the observed data period. 
\begin{table}[H]
    \centering
    \begin{tabular}{ccc}
         Parameter & Interpretation & Value\\
           \hline \\
           $1/\gamma$ & Mean latent period duration & 4\\
           $1/\nu$ & Mean infectious period duration & 7\\
           $1/\eta$ & Mean duration when recovered but still shedding RNA & 18 \\
           $\rho$ & scales concentrations of RNA into observed concentrations & $0.011$\\
           $\tau$ & describes the noisiness of observed gene data & 0.5 
\end{tabular}
\caption{Simulation parameters used in the simulation study. Durations are measured in days.}
\label{tbl:simsetting}
\end{table}
The estimate for $1/\gamma$ was calculated by averaging the mean latent period calculated by \citet{Xin2021} with the mean time to detecting virus which could be cultured found in \citep{killingley2022safety}. We took culturable virus to be a proxy for infectiousness.  
The mean time from infectiousness to symptom onset was 1.37, calculated again as an average from the previous two studies (1.4 days from \citep{Xin2021} versus 1.33 from \citep{killingley2022safety}). 
Mean infectious period was calculated using the mean period of detecting virus which could be cultured found in \citep{killingley2022safety}. 
Many studies have calculated the time from symptom onset to the end of RNA shedding in fecal matter; we averaged the estimates from \citep{okita2022duration} and \citep{zhang2021prevalence}. 
Because these two literature reviews shared studies, we dis-aggregated their estimates into individual study estimates, counting each study only once in our final average. We used mean shedding estimates from each study reported by \citet{okita2022duration}. 
\citet{zhang2021prevalence} did not include estimates of the mean for each study in their review, if estimates of the mean duration were available from the original paper, they were used, if they were not, the paper was not included in the final average. 
We also examined the literature review by \citet{walsh2020sars}, but found no new studies with more than two samples not in the previous two reviews. 
We decided to exclude studies with fewer than three samples. 
The final average duration from symptom onset to the end of RNA shedding in fecal matter was 22.99 days. 
We calculated $1/\eta$ (the mean duration of shedding after recovery) as 
\[
1/\eta = 22.99 + 1.33 - 1/\nu = 17.86.
\]
Note we are assuming shedding begins at the start of the infectious period. 
Note that all of the parameters are based on studies of the original Wuhan lineage of SARS-CoV-2. 
To calculate a plausible $\tau$ (standard deviation from true genetic concentration), we used the JWPCP data, and fit a Bayesian thin plate regression generalized student t-distribution spline to the data. We used the mean of the posterior estimate of $\tau$ for the $\tau$ in our simulation. 
The value of df was chosen as the mean of the posterior estimate from the thin plate spline model.
The value for $\rho$ (scaling factor for the mean of the generalized t-distribution) was chosen arbitrarily so that the mean total genome concentrations produced by the simulation would be roughly similar to the means seen in the JWPCP data. 
The value for $\phi$ (the negative binomial over-dispersion parameter) was chosen by fitting a negative-binomial spline to Los Angeles case data, and using the mean from the posterior estimate for $\phi$.
The value for $\psi$ (mean case detection rate) we chose based on our study of the spread of SARS-CoV-2 in Orange County, CA, the neighboring county to Los Angeles. \citep{bayer2023semi}. In that study, we estimated a weekly case detection rate, and at the end of the study period we estimated between 1 in 5 and 1 in 7 new infections were being detected. 
We chose to use 0.2 (1 in 5 cases being detected) for our simulation study.
\begin{table}[H]
\caption{Priors used by all models in the baseline simulation scenario.}
\centering
\small
\fbox{%
\begin{tabular}{*{5}{c}}
          Parameter & Model & Prior & Prior Median (95\% Interval) & Truth \\
         \hline \\
         $\gamma$ & All & Log-normal($\log(1/4)$, 0.2) & 0.25 (0.17, 0.37) & 0.25 \\
         $\nu$ &  All &  Log-normal($\log(1/7)$, 0.2) & 0.14 (0.10, 0.21) & 0.14  \\
         $\sigma_{rw}$ & All & Log-normal(log(0.1), 0.2) & 0.1 (0.07, 0.15) & NA \\
         $\eta$ & SEIRR-ww/EIRR-ww &  Log-normal($\log(1/18)$, 0.2 ) &  0.06 (0.04 0.08) & 0.06 \\
         $R_{0,0}$ &  SEIRR-ww/SEIR-cases & Log-Normal(log(0.88), 0.1) & 0.88 (0.72, 1.07) & 0.9 \\
         $\lambda$ & SEIRR-ww/EIRR-ww &  Logit-normal(5.69, 2.18) & 0.997 (0.81, 1) & NA \\
         $\tau$ & SEIRR-ww/EIRR-ww &  Log-normal(0, 1) & 1.00 (0.14, 7.10) & 0.5\\
         $df$ & SEIRR-ww/EIRR-ww & Gamma(10,2) & 19.33 (9.59, 34.17) & 2.99 \\
         $\rho$ & SEIRR-ww/EIRR-ww & Log-normal(0, 1)  & 1.00 (0.14, 7.10) & NA \\
         $R_{0}$ &  EIRR-ww/EIR-cases & Log-Normal(log(0.88, 0.1) & 0.88 (0.72, 1.07) & 0.80 \\
         $\psi$ & SEIR-cases/EIR-cases & Logit-Normal(-1.39, 0.4) & 0.20 (0.10, 0.35) & 0.2 \\
         $\phi$ & SEIR-cases/EIR-cases & Log-Normal(4.22, 0.29) & 68.03 (38.54, 120.11) & 57.55 \\
         $S\_SEIR1$ & SEIRR-ww & Logit-Normal(3.47, 0.05) & 0.97 (0.967, 0.972) &  0.97\\
         $I\_EIR1$ & SEIRR-ww &  Logit-Normal(-1.548302, 0.05) & 0.18 (0.16, 0.19) & 0.18  \\
         $R1\_ER1$ & SEIRR-ww &  Logit-Normal(2.22, 0.05) & 0.90 (0.89, 0.91) & 0.90 \\
         $S\_EI$ & SEIR-cases & Logit-Normal(4.83, 0.05) & 0.992 ( 0.991, 0.993) & 0.992 \\
         $I\_EI$ & SEIR-cases & Logit-Normal(0.78, 0.05) & 0.68 (0.66, 0.71) & 0.68 \\
         $E(0)$ & EIRR-ww/EIR-cases & Normal(225, 0.05) & 225.00 (224.90, 225.01) & 225 \\
         $I(0)$ & EIRR-ww/EIR-cases & Normal(489, 0.05) & 489.00 (448.90, 489.01) & 489 \\
         $R1(0)$ & EIRR-ww & Normal(2075, 0.05) & 2075.00 (2074.90, 2075.01) & 2075 \\
\end{tabular}}
\label{tab:ww_sim_priors}
\end{table}

\begin{figure}[H]
    \centering
    \includegraphics[width = \textwidth]{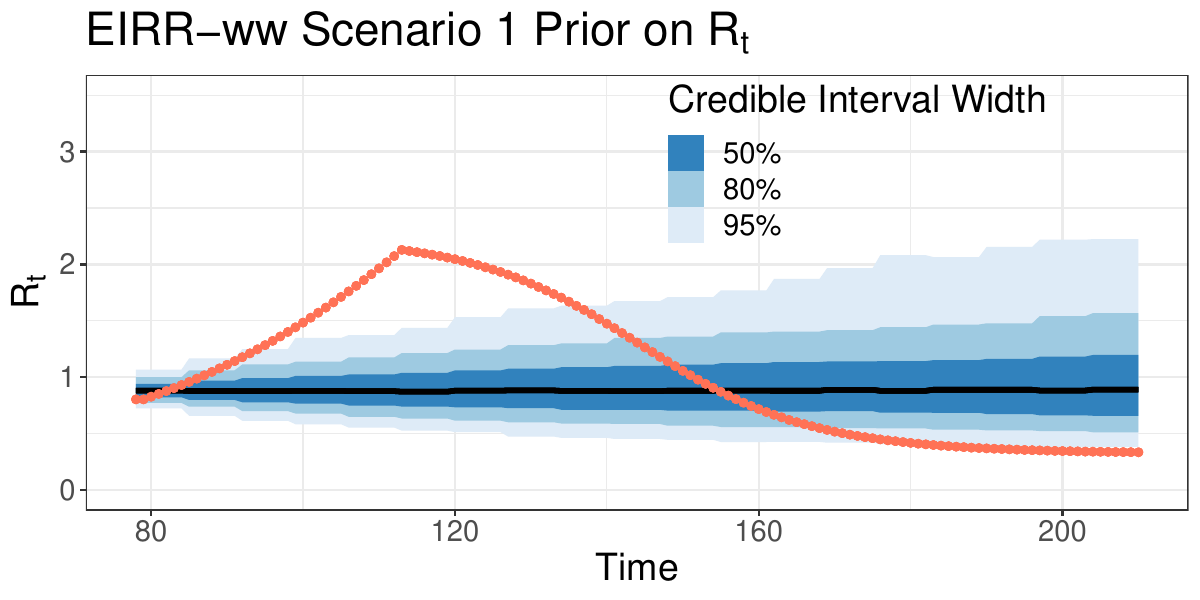}
    \caption{EIRR-ww prior summaries for the time-varying effective reproduction number. Blue bars are prior credible intervals, black lines are medians, true values are shown in red.}
    \label{fig:rt_prior}
\end{figure}

Figure \ref{fig:rt_prior} displays the prior quantiles of the random walk prior on $R_{t}$ for the baseline simulation scenario for the EIRR-ww model. 

\subsubsection{Realization of the Simulation}
\begin{figure}[H]
    \centering
    \includegraphics[width = 1.0\textwidth]{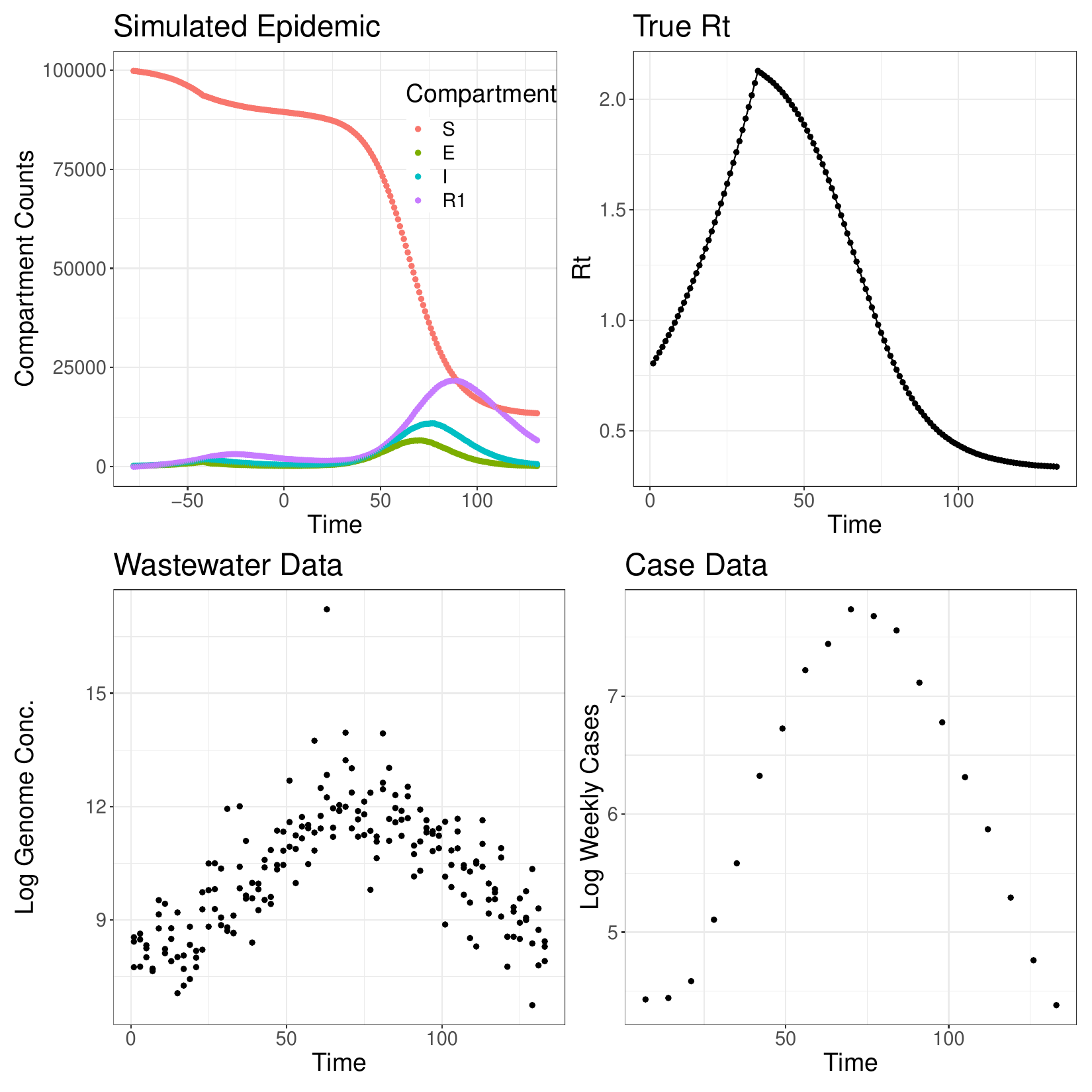}
    \caption{An example of simulated epidemic data generated from a single realization of a stochastic SEIRR model. Top left panel shows the simulated epidemic, time 0 is the day before the first wastewater sample/case is observed. Top right panel shows the true effective reproduction number trajectory. Bottom row shows simulated wastewater and case data on the log scale.}
    \label{fig:ww_simfig}
\end{figure}

\subsubsection{Computational Considerations}
The systems of ordinary differential equations for the EIR/EIRR models are linear, and thus can be solved in closed form. 
We used \texttt{Mathematica} Version (13.1) to calculate the closed form solution (see Appendix Section \ref{sec:closed_form}). 
We conducted informal tests on a Macbook Air M2 (2022) to compare performance of using the EIRR-ww model with the closed form solution we derived vs the Tsit5 solver \citep{tsitouras2011runge} implemented in the \texttt{Julia} package \texttt{DifferentialEquations} \citep{rackauckas2017differentialequations}. 
First, we compared the EIRR-ww model fit to one realization of the baseline simulation scenario using the closed form solution versus the EIRR-ww model using the ODE solver. 
Next, we created a new data set using the same dynamics as the baseline scenario, but where wastewater data was sampled every twelve hours, so that the solution to the ODE system had to be solved every half-day as opposed to every day.
The results of these experiments are displayed in Table \ref{tbl:comp_time}. 

\begin{table}[H]
    \centering
    \begin{tabular}{ccc}
         Setting & Closed Form Time & Numerical Solver Time \\
           \hline \\
           Fitting to day data & 855 seconds & 872 seconds\\
           Fitting to half-day data & 3567 seconds & 4260 seconds \\
    \end{tabular}
    \caption{Comparing performance of closed form solution to ODE system versus an ODE solver.}
    \label{tbl:comp_time}
\end{table}

Although the computational benefit of using the closed form ODE solution is not substantial, not having to solve ODEs numerically has other advantages.
Using an ODE solver inside an MCMC algorithm is a non-trivial undertaking, as tuning parameters controlling the tolerances for solver errors must be specified by the users \citep{timonen2022importance}. 
Using the closed form solution, our informal testing suggests our computation times are at least as good as the ODE solver, and regardless of computation time, our method is more robust as it does not require properly specifying error tolerance tuning parameters. 
\par

\subsubsection{EIRR-ww model performance in scenario 1}
\begin{figure}[H]
    \centering
    \includegraphics[width = \textwidth]{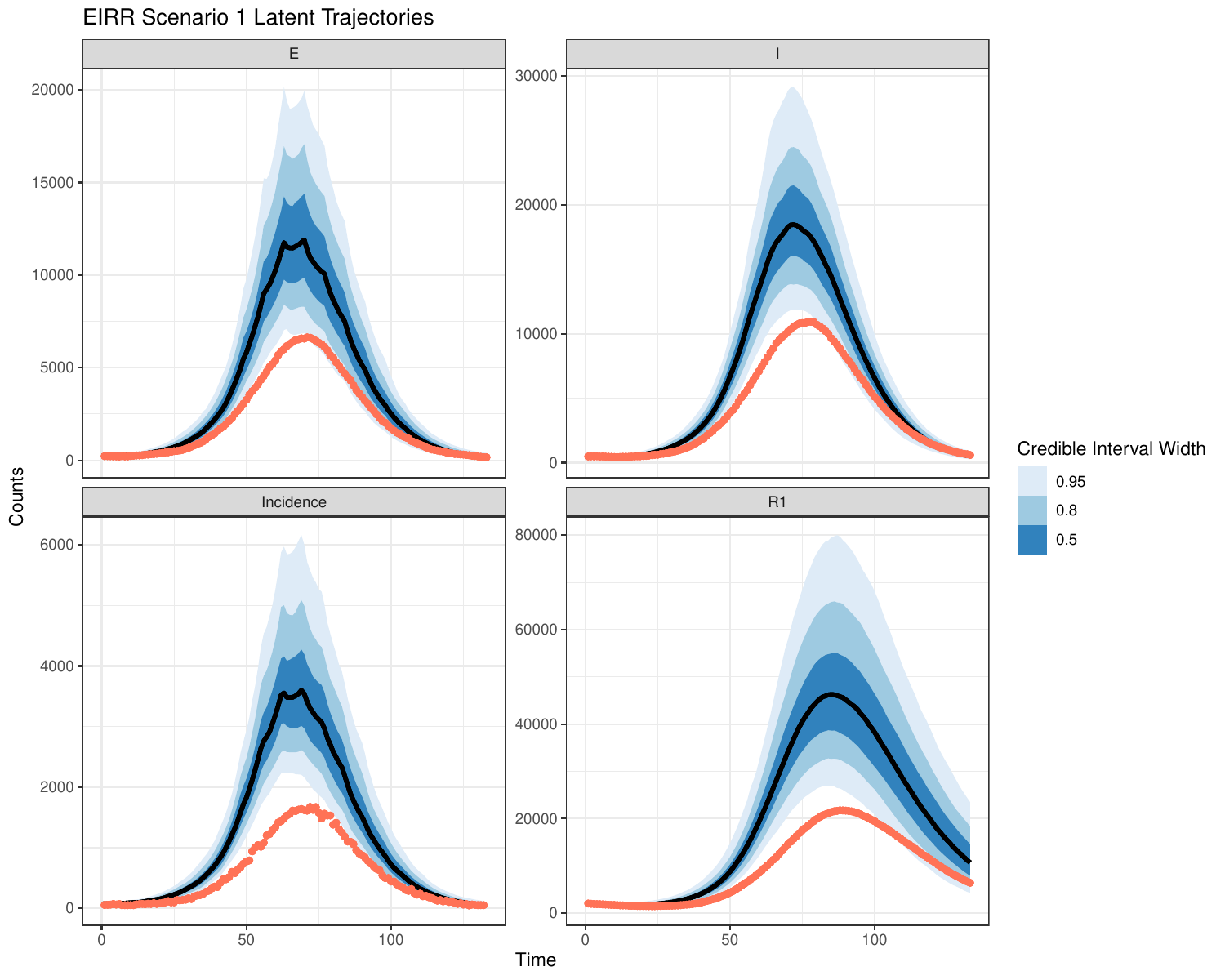}
    \caption{EIRR-ww posterior summaries for latent unobserved compartments and incidence. Posterior is from the model fit shown in Figure 1. Blue bars are credible intervals, black lines are medians, true values are shown in red. }
    \label{fig:s1_compartments}
\end{figure}

\begin{figure}[H]
    \centering
    \includegraphics[width = \textwidth]{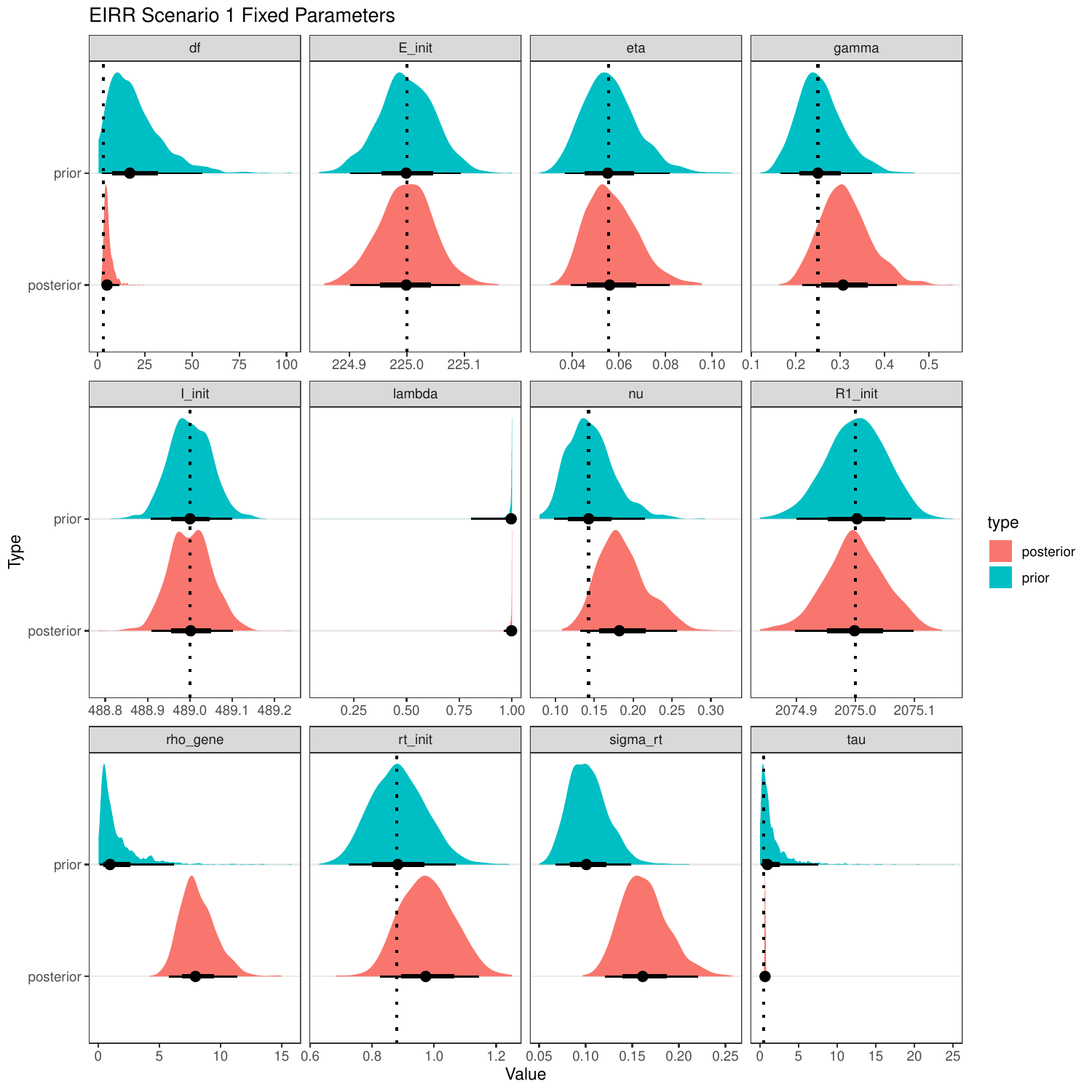}
    \caption{EIRR-ww prior and posterior summaries for fixed model parameters.  Posterior summaries are from the model fit shown in Figure 1. Blue densities are the prior, red densities are the posterior, dotted lines indicate true values (when relevant).}
    \label{fig:s1_fixed}
\end{figure}

\begin{figure}[H]
    \centering
    \includegraphics[width = \textwidth]{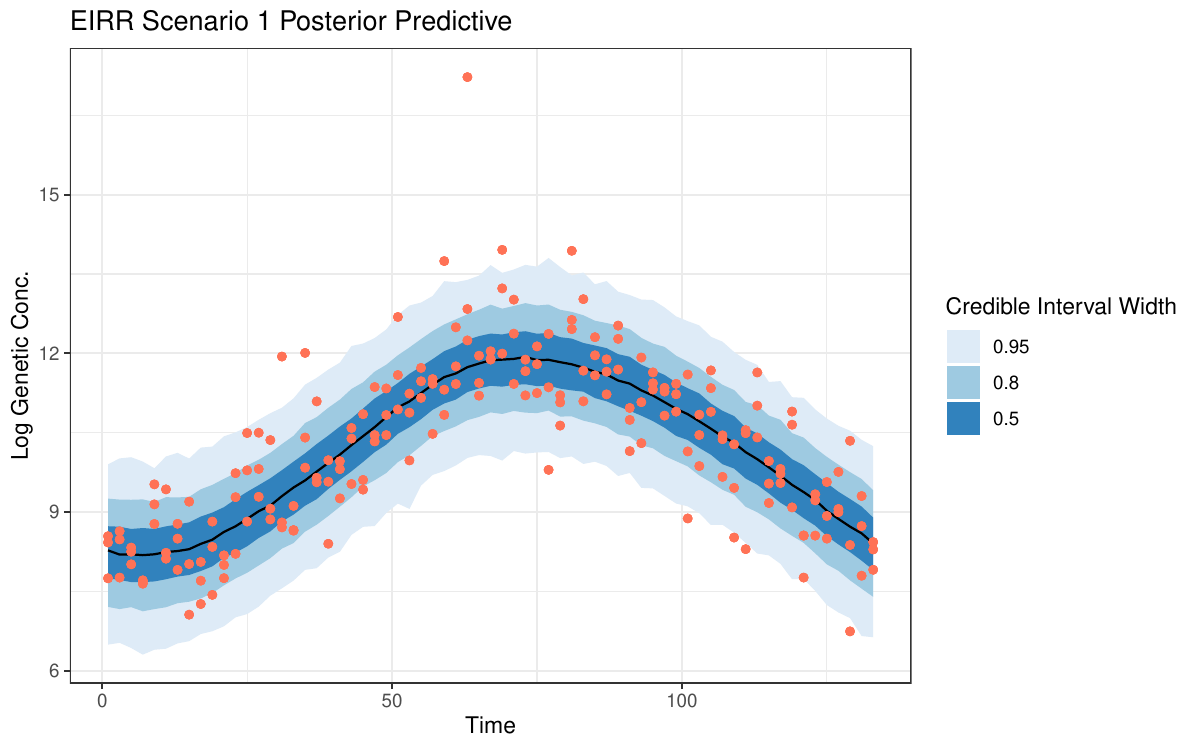}
    \caption{EIRR-ww posterior predictive of Log genome concentrations. Posterior predictive summaries are from the model fit shown in Figure 1. Blue bars represent credible intervals, black lines medians, and red dots are observed data.}
    \label{fig:my_label}
\end{figure}

\subsubsection{Model Performance Using Finer Random Walk Grid}
\begin{figure}[H]
    \centering
    \includegraphics[width = \textwidth]{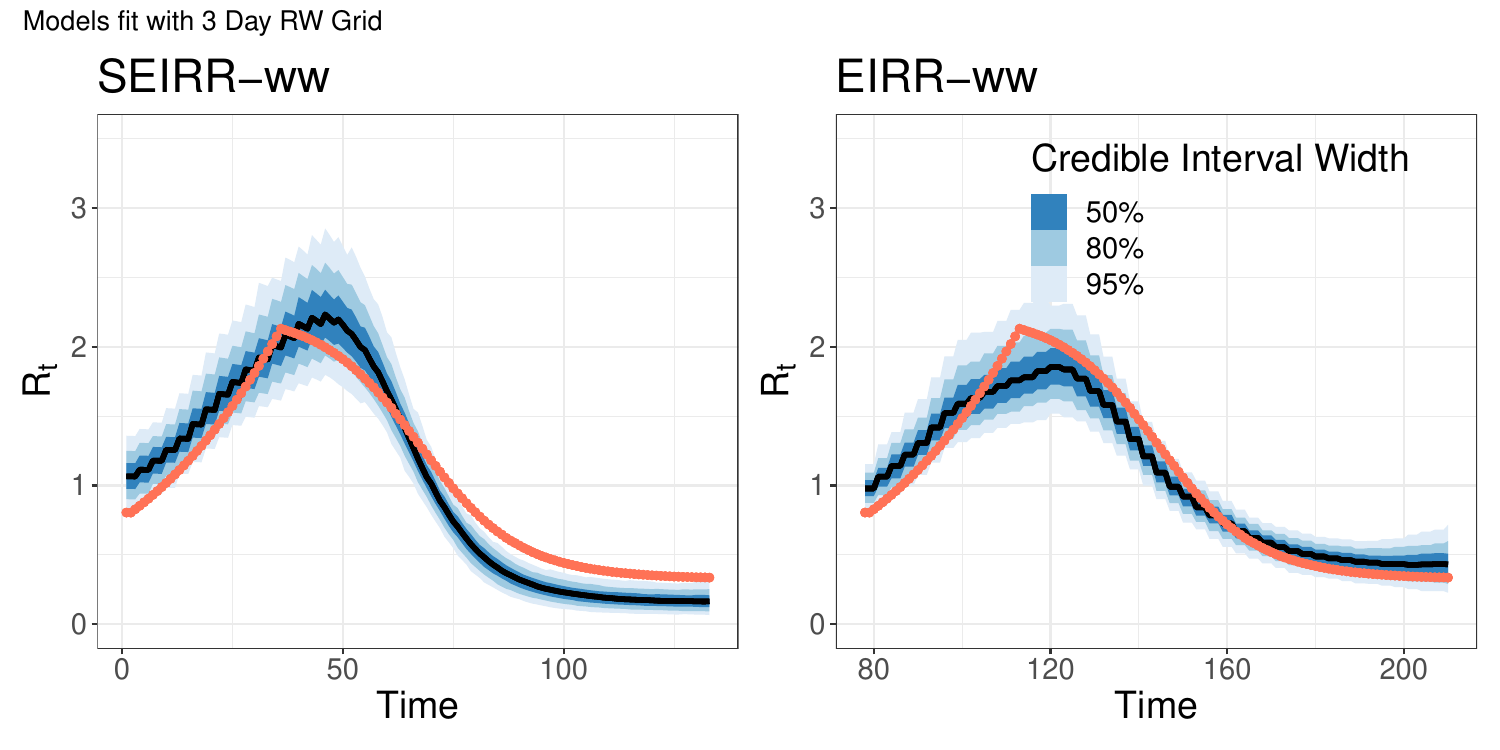}
    \caption{Posterior summaries of Rt using two models fit to wastewater data using a random walk grid size of 3 days. This is a recreation of Figure 1 in the main text using a grid size of 3 days for the random walk prior, as opposed to 7 days.}
    \label{fig:smallgrid_plot}

\end{figure}

We used a grid size of 7 days in our analyses, but finer grids are possible. 
Figure \ref{fig:smallgrid_plot} shows a recreation of Figure 1 in the main text using a finer grid size, the end result is smoother posterior estimates, and while there are some differences in inference between the two model fits, they are largely the same.
We chose a 7 day grid as it performs well in simulation, reduces computational burdens, and is more resistant to spurious changes caused by high variation in wastewater data.

\subsubsection{MCMC Diagnostics}
We used the \texttt{posterior} package to calculate $\hat{R}$, ESS bulk and ESS Tail for all parameters for our models in order to assess convergence of MCMC chains \cite{vehtari2021rank}.
We display the minimum and maximum across all simulations for each of our models and scenarios. 
We concluded the model had converged if $\hat{R}$ was below 1.05 and the ESS were both above 100. 
If this was not the case, we re-ran the model for an increased number of iterations (increasing from 500 iterations per chain to 1000 iterations per chain).
\begin{table}[H]
\centering
\footnotesize
\begin{tabular}{rlrrrrrr}
  \hline
 Scenario & Max Rhat & Min Rhat & Max ESS Bulk & Min ESS Bulk & Max ESS Tail & Min ESS Tail \\ 
  \hline
EIR-cases & 1.03 & 1.00 & 7748.92 & 514.43 & 4124.70 & 163.18 \\ 
   EIR-cases LA & 1.02 & 1.00 & 1213.06 & 259.83 & 1060.19 & 337.05 \\ 
  EIRR-ww (1) & 1.03 & 1.00 & 4452.76 & 649.23 & 2036.34 & 228.85 \\ 
  EIRR-ww (10 mean) & 1.03 & 1.00 & 2969.81 & 300.12 & 2057.70 & 114.07 \\ 
  EIRR-ww (10) & 1.03 & 1.00 & 2176.22 & 502.38 & 1133.52 & 234.43 \\ 
  EIRR-ww (3 mean) & 1.03 & 1.00 & 5071.60 & 406.84 & 2053.79 & 109.91 \\ 
  EIRR-ww (3) & 1.03 & 1.00 & 3631.31 & 528.05 & 2149.23 & 226.18 \\ 
  EIRR-ww High Init & 1.03 & 1.00 & 3342.74 & 622.39 & 2042.90 & 290.15 \\ 
  EIRR-ww LA & 1.02 & 1.00 & 2249.27 & 842.62 & 1129.10 & 375.09 \\ 
   EIRR-ww Low Init & 1.03 & 1.00 & 4467.27 & 425.32 & 2151.38 & 126.54 \\ 
  EIRR-ww Low Prop & 1.03 & 1.00 & 2717.95 & 436.85 & 1132.62 & 100.07 \\ 
  EIRR-ww Stoch Rt & 1.03 & 1.00 & 3000.00 & 538.30 & 1135.72 & 217.10 \\ 
  SEIR-cases & 1.03 & 1.00 & 4145.55 & 414.77 & 2059.80 & 167.21 \\ 
  SEIRR-ww & 1.03 & 1.00 & 4582.79 & 463.60 & 2101.94 & 142.96 \\ 
   \hline
\end{tabular}
\end{table}

\subsubsection{Additional Simulation Results}
\begin{figure}[H]
    \centering
    \includegraphics[width = \textwidth]{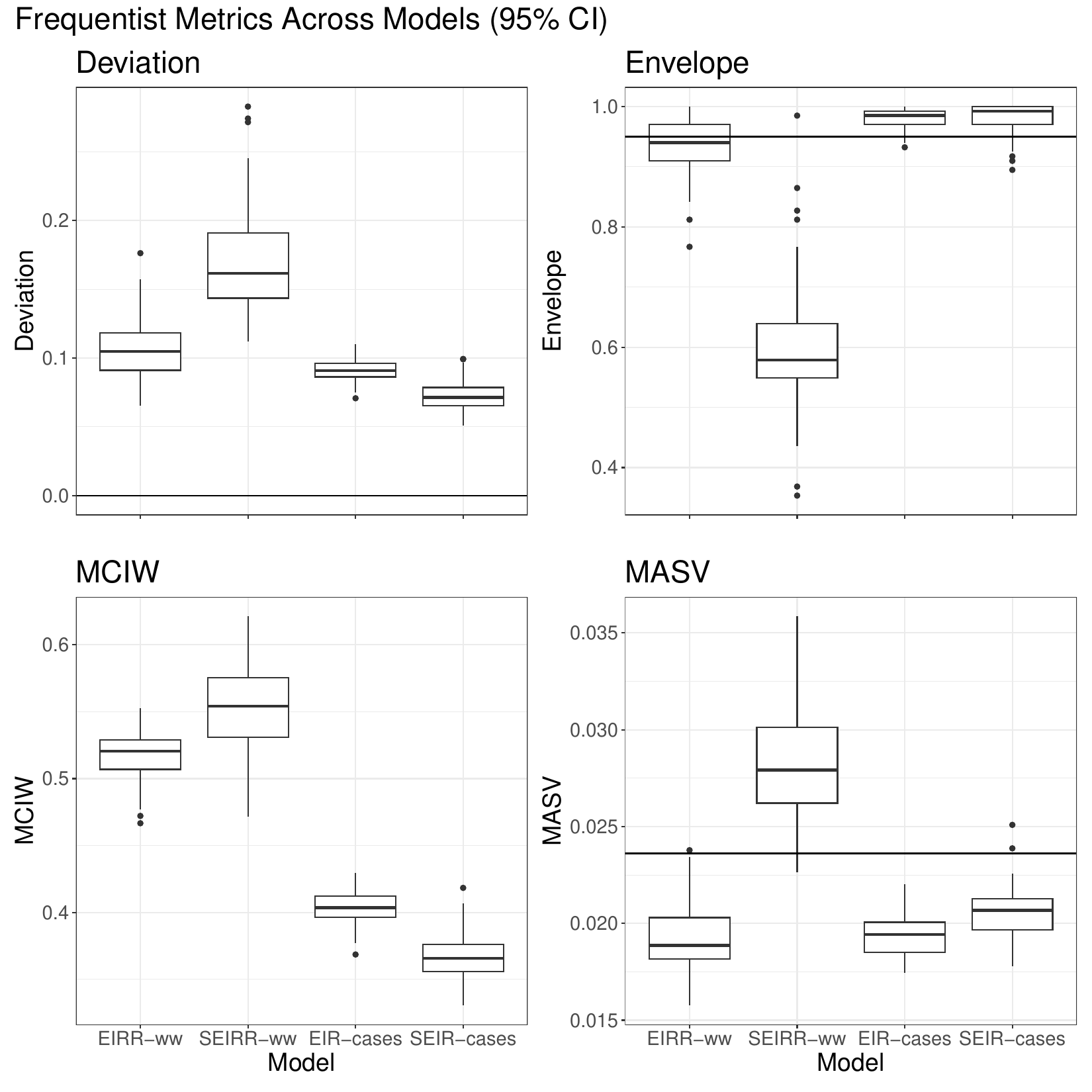}
    \caption{Frequentist metrics for all models using 95\% CI as opposed to 80\% CI as in the main text. The ideal Envelope is now 0.95. See Figure 2 in the main text for descriptions of the metrics.}
    \label{fig:freq_metrics_95CI}

\end{figure}

\begin{figure}
    \centering
    \includegraphics[width = 1.0\textwidth]{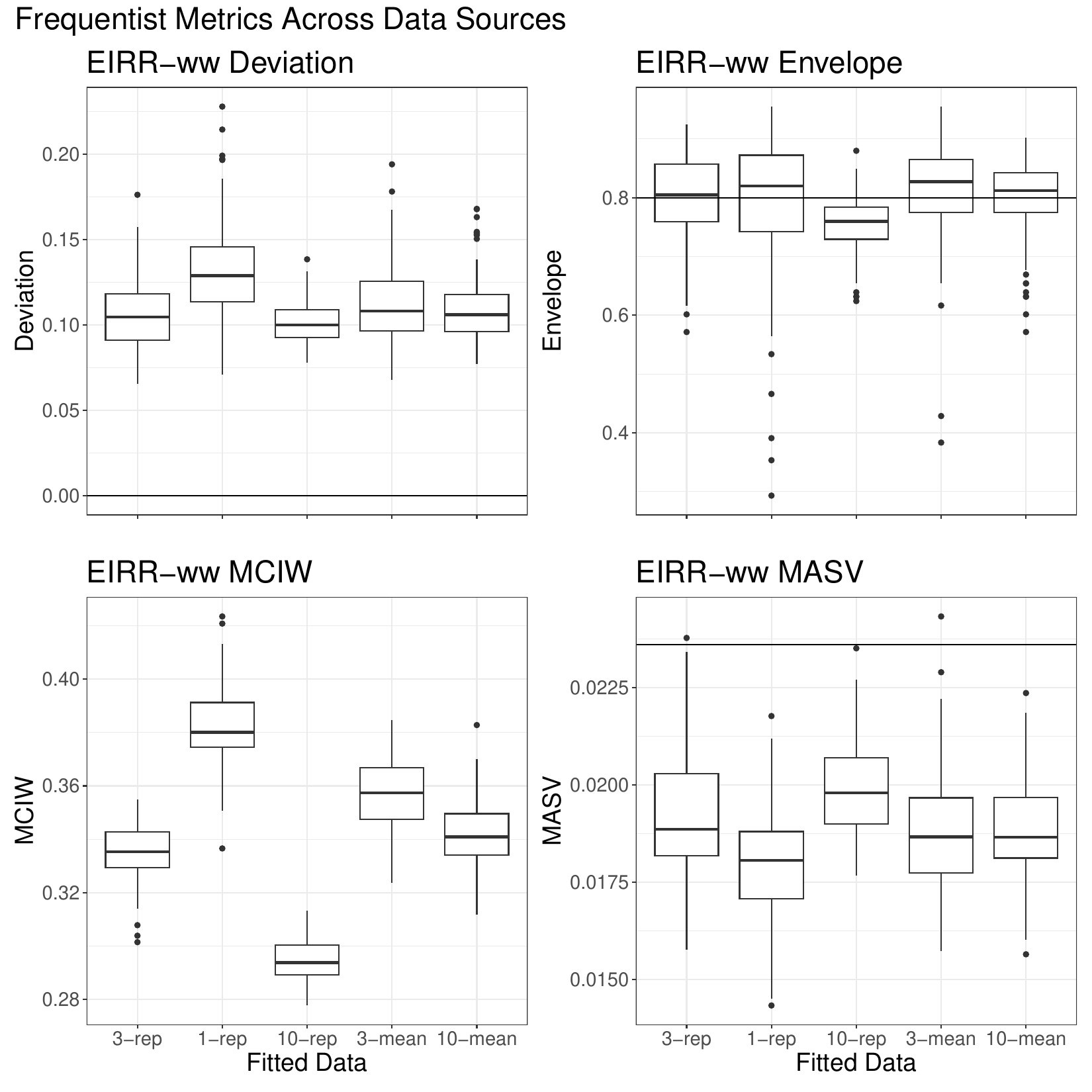}
    \caption{Frequentist metrics for performance of the EIRR-ww model using different kinds of wastewater data. The baseline scenario 3-reps uses three replicates,  1-rep uses one replicate, 10-reps uses ten replicates, 3-mean uses the mean of three replicates, 10-mean uses the mean of ten replicates. See Figure 2 for descriptions of the metrics.}
    \label{fig:freq_metrics_data}
\end{figure}

\begin{figure}[H]
    \centering
    \includegraphics[width = \textwidth]{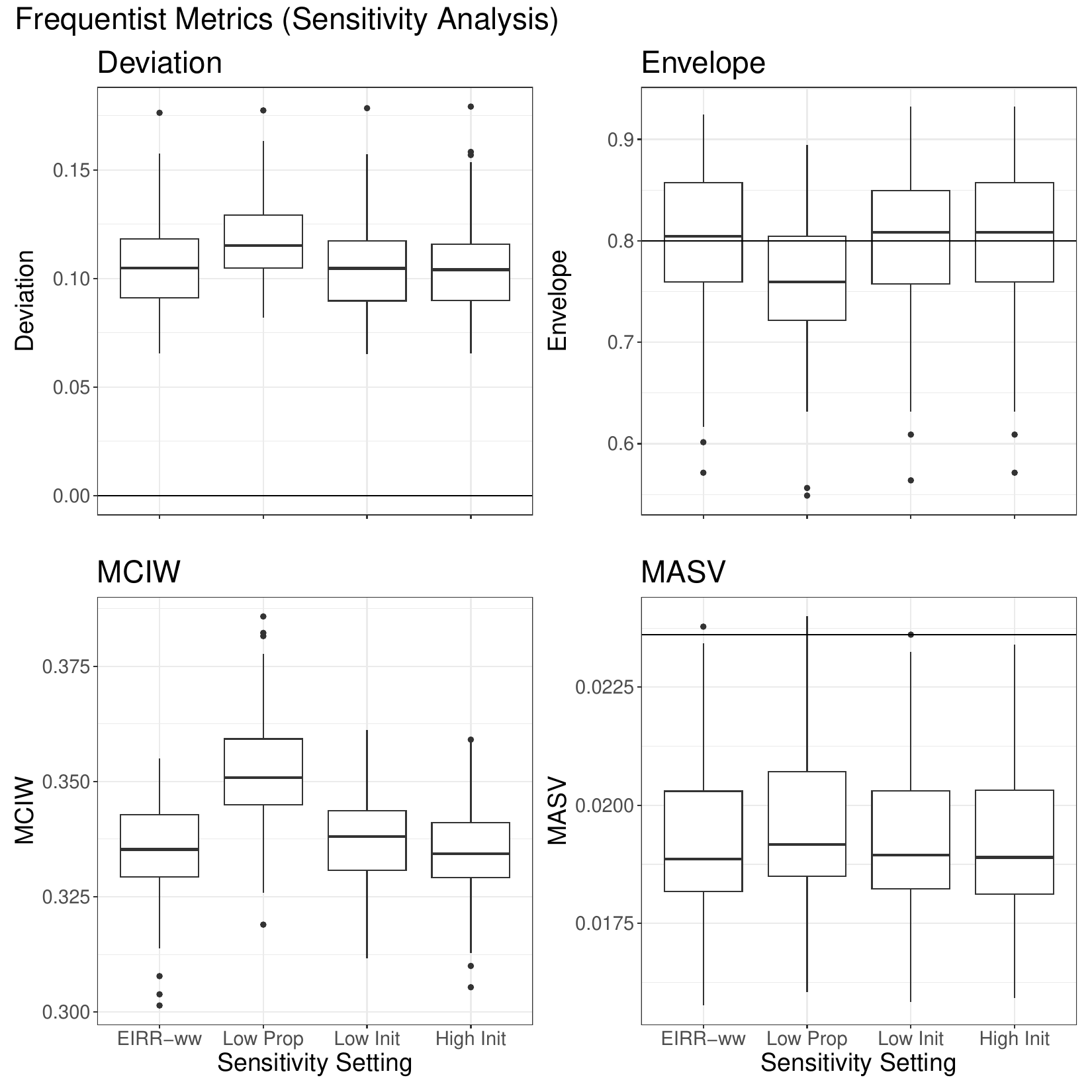}
    \caption{Frequentist metrics for performance of the EIRR-ww model using mis-specified priors. See Figure 2 in the main text for descriptions of the metrics.}
    \label{fig:freq_metrics_sensitivity}
\end{figure}

Note that for the Huisman method, because the method relies on generating a synthetic series of cases, the method truncates the inferred $R_{t}$ values in order to only infer $R_{t}$ for times for which the method believes there are enough pathogen genome concentrations available to correctly generate synthetic cases. 
Also, \texttt{EpiEstim} does not estimate the effective reproduction number for early points in the time-series.
When comparing the Huisman method to the baseline EIRR-ww model, we restrict the comparison to only be on time points for which both models produced inference.

\begin{figure}[H]
    \centering
    \includegraphics[width = \textwidth]{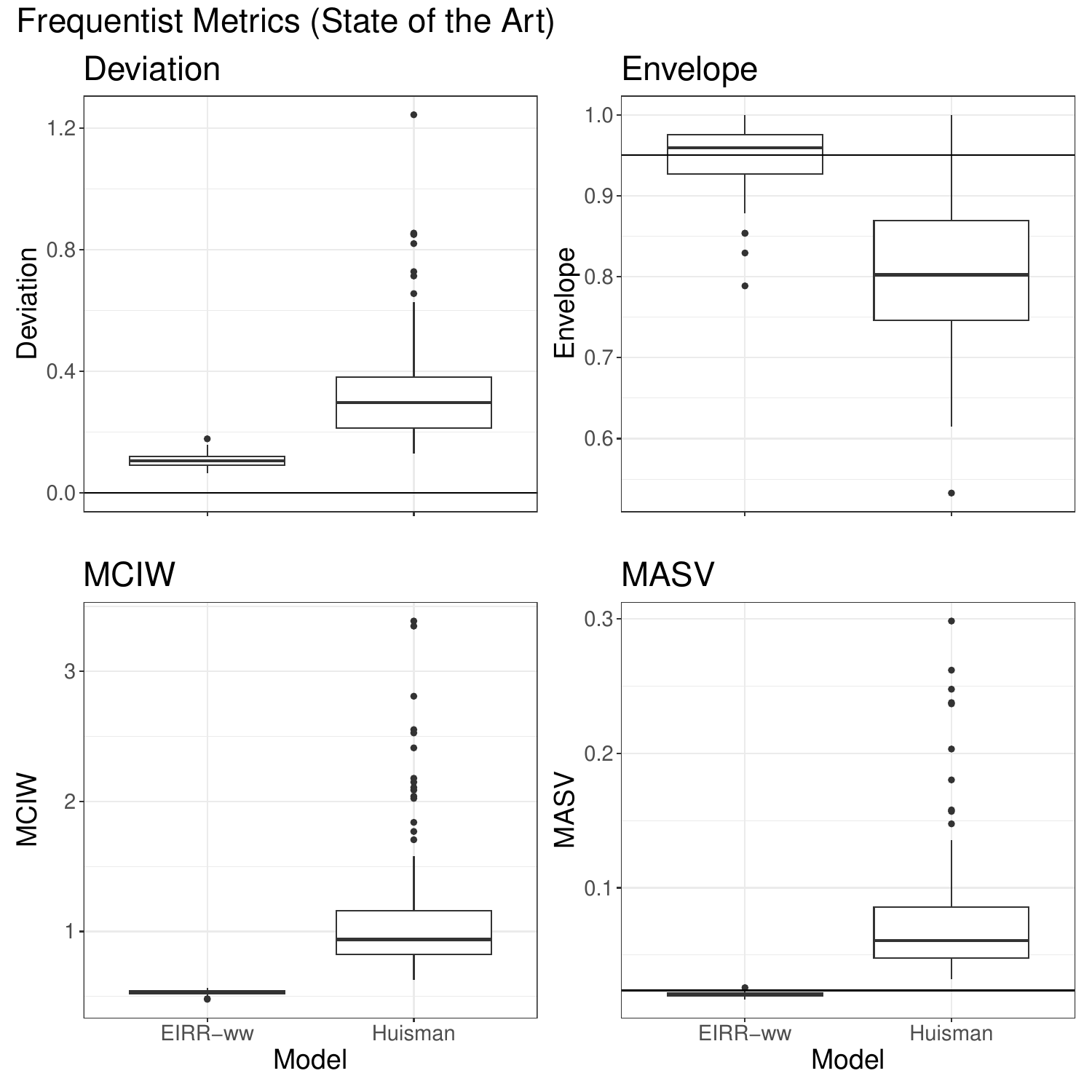}
    \caption{Frequentist metrics for performance of the EIRR-ww model compared to the Huisman et al. (2022) method.
    See Figure 2 in the main text for descriptions of the metrics.
    Four simulations were removed from the figure because the Huisman model had unusual large metric values, making visual comparisons difficult.
    Simulations were removed if the Huisman model had either a deviation over 600, and MCIW over 5 or MASV above 2. 
}
    \label{fig:freq_metrics_huisman}
\end{figure}

\begin{figure}[H]
    \centering
\includegraphics[width = \textwidth]{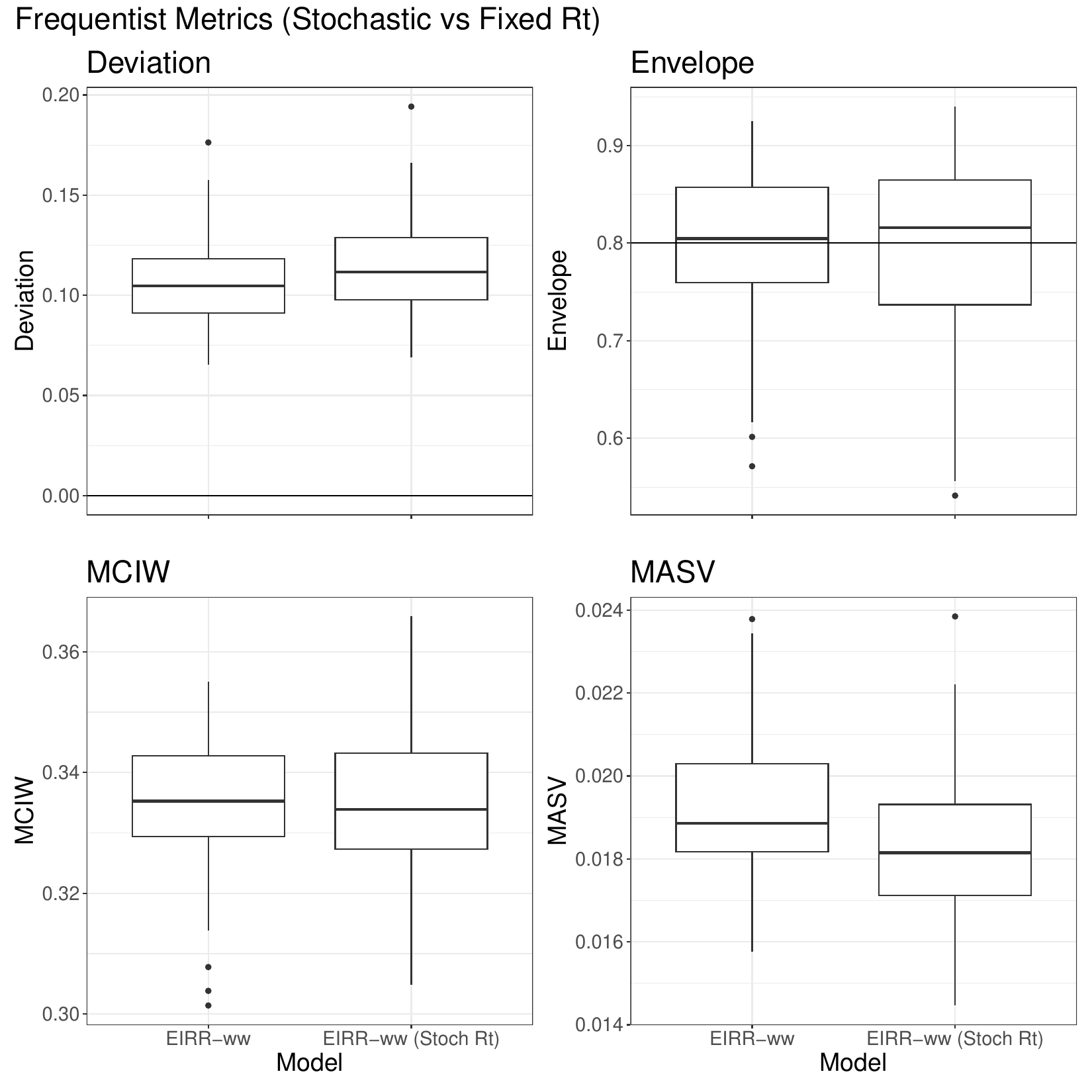}
    \caption{Frequentistic metrics for performance of the EIRR-ww model on 100 simulations where $R_{t}$ is also simulated. In the baseline scenario, $R_{t}$ is fixed for all simulations, we compare to the case where $R_{0}$ is fixed, but $R_{t}$ varies. See Figure 2 in the main text for descriptions of the metrics.}
    \label{fig:enter-label}
\end{figure}

\subsubsection{Calculating Initial Conditions for Los Angeles, CA}\label{la_init_conds}
We used the case data to create a rough guess for the initial conditions for our EIRR-ww and EIR-cases models. 
We assumed the total number of individuals in Los Angeles County in the $E$ and $I$ compartments was equal to the last 11 days before the start of the observation period (the sum of the average latent period and average infectious period) of reported cases multiplied by 5 (i.e. an under-reporting rate of 0.2). We then split this total so that two thirds went to the $I$ compartment and one third went to the $E$ compartment. 
We then took the 18 days (the average duration in the $R1$ compartment) before the last 11 days, multiplied the total cases by 5 and assumed that this was the number of individuals in the $R1$ compartment. 
Los Angeles County has about 10 million people total, so we multiplied these counts by 0.48 for the final compartmnet counts, as the JWPCP plant serves 4.8 million people. 
For the initial effective reproduction number, we chose a prior centered around 2, this was based on previous estimates of the effective reproduction number during this time using case data \citep{goldstein2022incorporating}.
The final priors are displayed in the table below. 
\begin{table}[H]
\caption{EIRR-ww/EIR-cases Priors for Los Angeles, CA.}
\centering
\fbox{%
\begin{tabular}{*{5}{c}}
          Parameter &  Prior & Prior Median (95\% Interval) \\
         \hline \\
         $E(0)$ & Normal(2995, 0.05) & 2995.00 (2994.90, 2995.10) \\
         $I(0)$ &  Normal(5990, 0.05) & 5990.00 (5998.90, 5990.10)  \\
         $R1(0)$ &  Normal(11055, 0.05) & 11055.00 (11054.90, 11055.10)\\
         $R_{0}$ & Log-Normal(log(2), 0.1) & 2 (1.64, 2.43)
\end{tabular}}
\label{tab:la_init_conds}
\end{table}

\subsubsection{Additional estimates of $R_{t}$ for SARS-CoV-2 in Los Angeles, CA}
\begin{figure}[H]
    \centering
    \includegraphics[width = \textwidth]{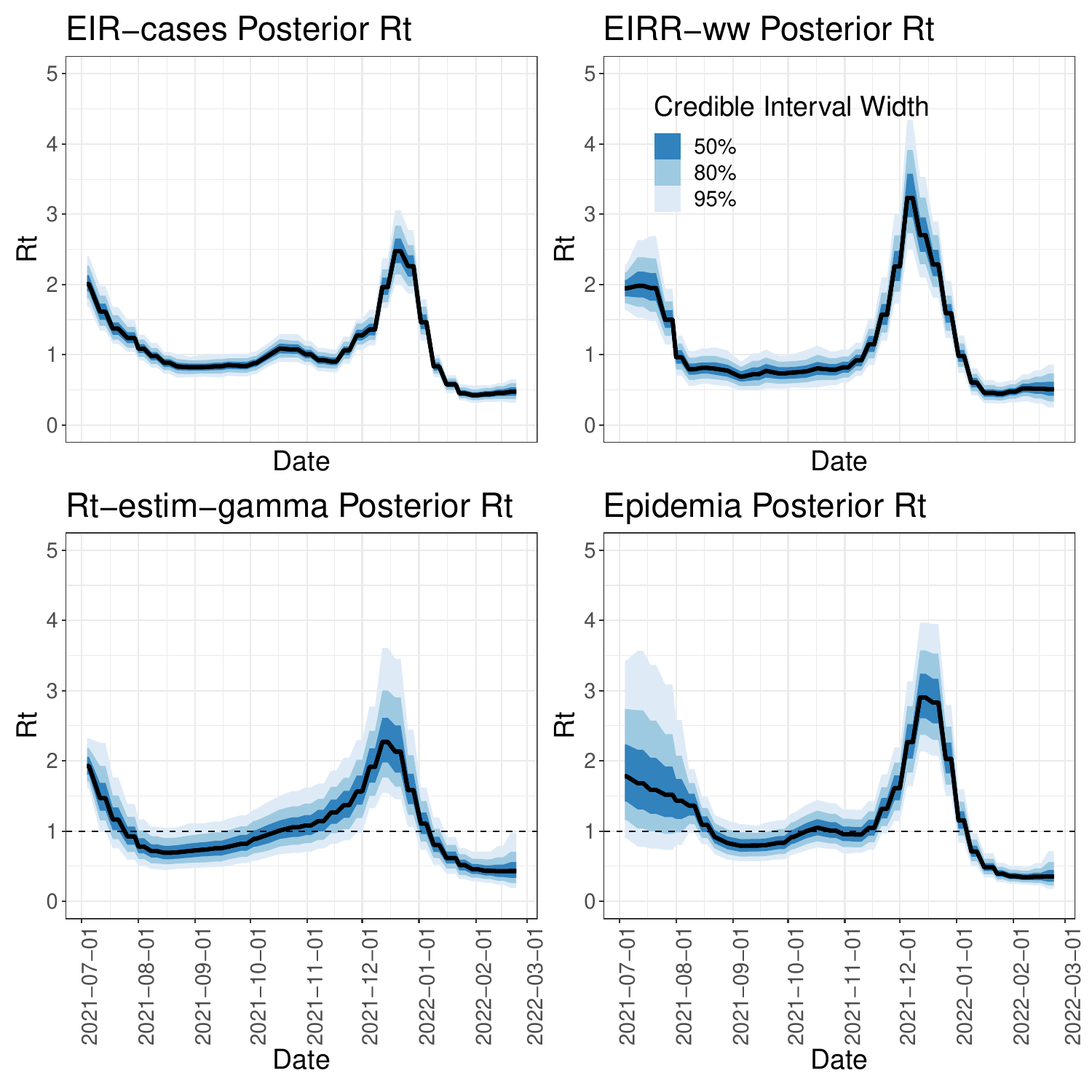}
    \caption{Posterior estimates of the effective reproduction number for the SARS-CoV-2 epidemic in Los Angeles, CA. Blue bars from dark to light represent 50, 80, and 95\% credible intervals. Black lines represent posterior medians. EIRR-ww model is fit to wastewater data. EIR-cases and epidemia are fit to cases alone. Rt-estim-gamma is fit to cases and uses total diagnostic tests as a covariate.}
    \label{fig:app_la_rt}
\end{figure}

\subsubsection{Estimates of the case detection rate for SARS-CoV-2 in Los Angeles, CA}\label{la_case_detection}
We can use the the posterior samples of $C(t_{u})$ (the cumulative incidence at time $t_{u}$) produced by the EIRR-ww model to estimate the case detection rate, i.e., the proportion of new infections which are observed as reported cases.
For each posterior sample, we calculate a sample of the case detection rate as 
\begin{equation*}
    \kappa_{u} = O_{u}/(C(t_{u} - C(t_{u-1}),
\end{equation*}
where $O_{u}$ is the number of new cases observed in the time period $(t_{u-1}, t_{u}]$. 
Note that $\kappa_{u}$ is a function of the total diagnostic tests administered during the time period, thus, it is also interesting to look at the case detection rate normalized by the number of diagnostic tests. We call this normalized case detection rate $\epsilon$, defined as 
\begin{equation*}
    \epsilon_{u} = \kappa_{u} / D_{u},
\end{equation*}
where $D_{u}$ is the total number of diagnostic tests (both positive and negative) administered in the time period $(t_{u-1}, t_{u}]$. 
Posterior estimates of the case detection rate and normalized case detection rate of SARS-CoV-2 from Los Angeles, CA are shown in the figure below.
\begin{figure}[H]
    \centering
    \includegraphics[width = \textwidth]{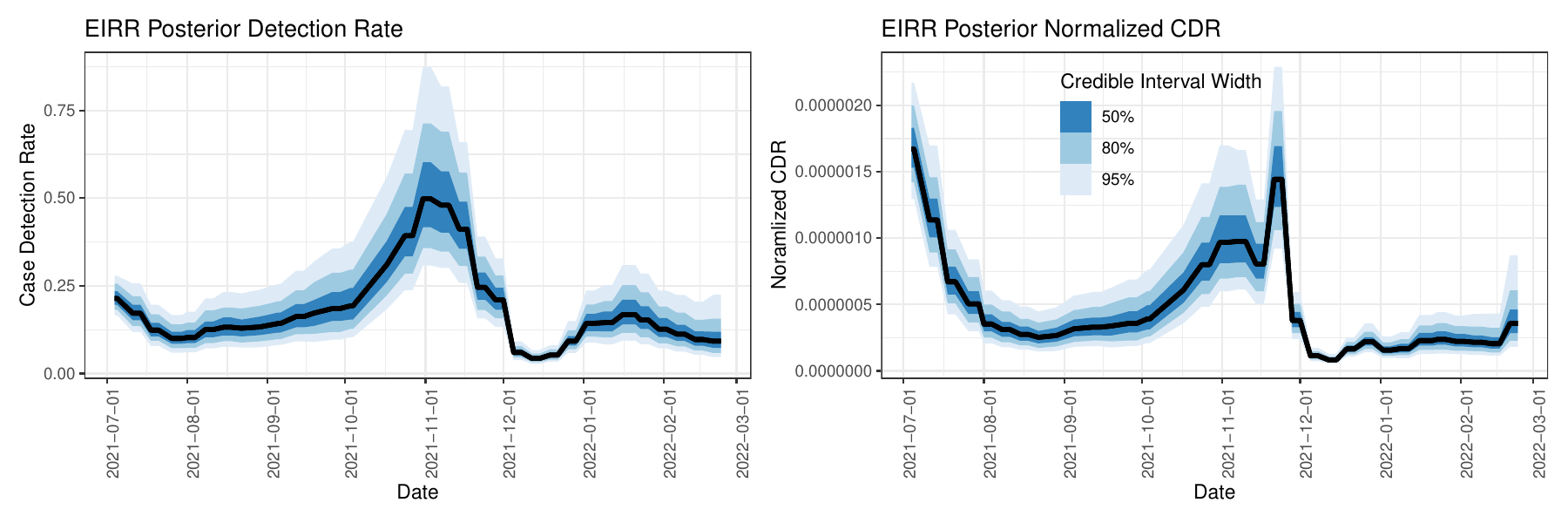}
    \caption{Posterior summaries of the case detection rate and case detection rate normalized by total diagnostic tests for SARS-CoV-2 in Los Angeles, CA. Posterior summaries of the number of new infections taken from wastewater, and the observed weekly case counts, are used to create posterior summaries of the case detection rate.}
    \label{fig:case_detect}
\end{figure}

\begin{figure}
    \centering
    \includegraphics[width = \textwidth]{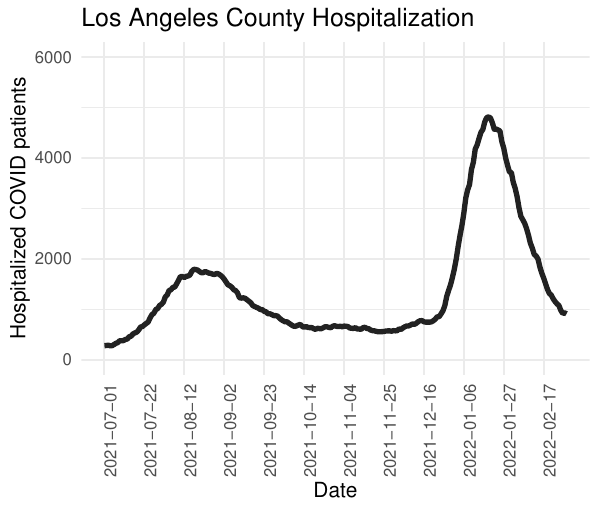}
    \caption{Time series of patients hospitalized with SARS-CoV-2 for Los Angeles County, CA.}
    \label{fig:la_hosp}
\end{figure}

\subsection{Discussion}
\subsubsection{Discussion of disagreements between case and wastewater models}
An interesting point of disagreement is in the period from October 2021 to November 2021, when all three case models have a posterior median around one by the middle of October. 
In contrast, the EIRR-ww posterior median is well below one. 
One explanation for the discrepancy is that the case detection rate may have changed in October 2021. 
To explore this possibility, we estimated the case detection rate using posterior estimates of the incidence from the EIRR-ww model (Figure \ref{fig:case_detect}), which indeed show a sharp increase in the case detection rate (both the raw rate and the rate normalized by the total diagnostic tests) in October. 
These estimates should be viewed skeptically, as we found that in simulation, the EIRR-ww estimates for incidence were not particularly accurate (Figure \ref{fig:s1_compartments}). 
Still, we would expect the EIRR-ww model to perform worst at peaks, when the linearity assumption is particularly violated (the same counts of individuals shed different amounts of genomes if they are recently infected versus near the end of infectiousness, leading to different concentrations), so the change shown in our model estimates may still be real. 
When we examined the time series of hospitalizations from SARS-CoV-2 during this time period (Figure \ref{fig:la_hosp}), a much more reliable data source than cases, it showed no sign of an increase in transmission rate in October (which we would expect to see reflected in an increase in hospitalizations in late October/early November). 
The other major point of disagreement is in July 2021, when the EIRR-ww model estimates $R_{t}$ to be stable around $2$ until August, while every case model estimates a steady decline in $R_{t}$. 
Here again the change in case detection rate may be to blame, but the hospitalizations are a little harder to interpret. 
The flatter hospitalization curve in August of 2021 may indicate the EIRR-ww model is correct, on the other hand, if $R_{t}$ was indeed still at $2$ on August 1st, we might expect a peak of hospitalizations even later than mid-August. 
We are inclined to think the case models are more correct in this case. 
\end{document}